\documentclass[12pt]{iopart}
\usepackage{graphicx}
\usepackage{dcolumn}
\usepackage{amssymb}
\usepackage{bm}
\usepackage{color}
\usepackage{cite}
\usepackage{hyperref}%
\hypersetup{
    colorlinks=true,
    linkcolor=blue,
    citecolor=blue,   
    urlcolor=blue,
}

\def\bea{\begin{eqnarray}}
\def\eea{\end{eqnarray}}
\def\beq{\begin{equation}}
\def\eeq{\end{equation}}
\def\f{\frac}

\def\t{\tau}

\def\la{\langle}
\def\ra{\rangle}
\def\nn{\nonumber}

\def\S{\Sigma}

\def\d{\delta}
\def\p{\partial}

\def\uv{ {\bf \hat{u}}}
\def\rv{ {\bf r}}

\def\d{\delta}
\def\p{\partial} 

\def\la{\langle}
\def\ra{\rangle}

\def\rpara{r_{\parallel}}
\def\rperp{\rv_{\perp}}

\def\Pe{\mathrm{Pe}}
\def\rpara{r_{\parallel}}
\def\rperp{r_{\perp}}

\begin{document}

\title[ABP under stochastic position and orientation resetting in a harmonic trap]{Active Brownian particle under stochastic position and orientation resetting in a harmonic trap}

\author{Amir Shee$^{1,2,3}$}
\address{$^1$Department of Physics, University of Vermont, Burlington, VT 05405, USA}
\address{$^2$Northwestern Institute on Complex Systems, Northwestern University, Evanston, IL 60201, USA}
\address{$^3$Department of Engineering Sciences and Applied Mathematics, Northwestern University, Evanston, IL 60208, USA}

\eads{\mailto{amir.shee@uvm.edu}}

\begin{abstract}

We present an exact analytical study of an active Brownian particle (ABP) subject to both position and orientation stochastic resetting in a two dimensional harmonic trap. Utilizing a Fokker-Planck-based renewal approach, we derive the system's exact moments, including the mean parallel displacement, mean squared displacement (MSD), and the fourth order moment of displacement, and compare these with numerical simulations. To capture deviations from Gaussian behavior, we analyze the excess kurtosis, which reveals rich dynamical crossovers over time. These transitions span from Gaussian behavior (zero excess kurtosis) to two distinct non-Gaussian regimes: an activity-dominated regime (negative excess kurtosis) and a resetting-dominated regime (positive excess kurtosis). Furthermore, we quantify the steady state phase diagrams by varying three key control parameters: activity, resetting rate, and harmonic trap strength, using steady state excess kurtosis as the primary metric.

\end{abstract}
%


\section{Introduction}
\label{sec_int}

Active particles, by converting energy into motion, exhibit a broad spectrum of dynamical behaviors~\cite{Romanczuk2012, Marchetti2013, Cates2015, Bechinger2016, Ramaswamy2017, Baconnier2024}. These behaviors are observed across diverse systems, including cytoskeletal filaments in motor protein assays~\cite{Schaller2010, Sumino2012, Shee2021, Karan2023}, bird flocks~\cite{Ballerini2008}, fish schools~\cite{Katz2011}, and artificial systems like active colloids~\cite{Ebbens2016} and robots~\cite{Dauchot2019}. Active particles are inherently out of equilibrium, displaying collective phenomena such as flocking~\cite{Vicsek1995, Toner2005, Kumar2014}, clustering~\cite{Fily2012, Palacci2013, Slowman2016}, and activity-induced phase separation~\cite{Schwarz-Linek2012, Bhattacherjee2019}. Even at the single particle level, they exhibit diverse dynamical behaviors~\cite{Caprini2022}, including short or intermediate time ballistic motion~\cite{Basu2018, Shee2020, Majumdar2020, Santra2021, Shee2022, SheePRE2022, Patel2023, Pattanayak2024}, nonequilibrium steady states in confinement~\cite{Pototsky2012, Solon2015, Malakar2020, Chaudhuri2021, Caraglio2022, Caprini2023, Nakul2023, Patel2024exact}, and dynamical transitions in relaxation and first passage properties~\cite{Malakar2018, Basu2019, Dhar2019, Singh2019}.
Recent studies have demonstrated enhanced control over active agents for decision making by space dependent rotational diffusion coefficient~\cite{Fernandez-Rodriguez2020}, as well as using external or internal cues, such as magnetic fields~\cite{Beppu2024}, chemical gradients~\cite{Ziepke2022}, and acoustic waves~\cite{Deng2023, Zhang2023b}, to control their speed and orientation, can also extended to stochastic resetting~\cite{Evans2018, Santra2020JSTAT, Kumar2020, Baouche2024}.

Stochastic resetting is a protocol that resets a dynamical system to a specific state, primarily governing the nonequilibrium steady state (NESS) through a continuous influx of probability from resetting. This process can lead to dynamical transitions during the system's relaxation to the NESS and often results in a non-monotonic mean first passage time~\cite{EvansPRL2011, Evans2011JPA, Reuveni2016, Majumdar2015, Pal2017PRL}. The resetting strategy has broad applications across various systems, including population dynamics and biological processes, particularly in optimizing search problems~\cite{Manrubia1999, Montanari2002, Kussell2005, Kussell2005Genetics, Visco2010, Roldan2016, Evans2020JPA, Pal2022JPA}. The impact of resetting on different diffusive dynamics has also been extensively studied~\cite{Evans2013JPA, Whitehouse2013, Evans2014JPA, Pal2015, Mendez2016, MasoPuigdellosas2019, Pal2019JPA, Gupta2019JSTAT, Ahmad2019, Tal-Friedman2020, Singh2020, Mercado-Vasquez2020, Gupta2019JSTAT}. Resetting has been shown to be an optimal navigation strategy for target search, particularly in Brownian motion~\cite{EvansPRL2011} and under external potentials~\cite{Ray2020}, and it could play a crucial role in active systems.

The resetting of a free active Brownian agent has been studied recently in~\cite{Kumar2020, Sar2023}, with extensions to external confining and absorbing potentials in~\cite{Abdoli2021, Zhang2023, Gueneau2024}. Activity introduces new dynamical behavior due to the competition between the internal active timescale and the resetting timescale, first studied in~\cite{Evans2018}. The first passage properties of active Brownian particle (ABP) and run-and-tumble particle (RTP) under various resetting mechanisms have been explored in~\cite{Bressloff2020, Santra2020JSTAT, Schumm2021, Olsen2023}. The dynamics of an ABP under position stochastic resetting, orientation stochastic resetting, and combined position and orientation stochastic resetting, focusing on marginal probability distributions, were examined in Kumar et al.~\cite{Kumar2020}. The dynamics under orientation stochastic resetting, using the intermediate scattering function, were studied in~\cite{Baouche2024}. However, the exact dynamics of an ABP under both position and orientation resetting in a harmonic trap, along with an analysis of the steady state properties, remain unexplored. From the lens of active matter, it is crucial to delve into the intricate interplay between stochastic resetting and activity under the influence of an external force.

In this work, we investigate the exact dynamics of active Brownian particle (ABP) under complete stochastic resetting, involving both position and orientation, in a harmonic trap in two dimensions. We apply the method for exact moment calculations of stiff chains, as described by Hermans et al.\cite{Hermans1952}, to compute the exact moments for the active dynamics first shown in Shee et al.~\cite{Shee2020}. Using a Fokker-Planck-based approach, we derive the dynamical moment generating equation, which we then use to calculate all dynamical moments. This method has been previously applied to study ABP in a harmonic trap in Chaudhuri et al.~\cite{Chaudhuri2021}, with anisotropic translational noise~\cite{SheePRE2022}, fluctuating speed~\cite{Shee2022}, and chirality/torque~\cite{Pattanayak2024}, and has also been extended to inertial ABP~\cite{Patel2023, Patel2024exact}. Here, we focus on exact moment calculations of ABP under complete stochastic resetting in a harmonic trap using a renewal equation following the Fokker-Planck-based moments calculation. Moreover, we characterize the steady state behavior across various limits and parameter ranges using excess kurtosis.

The remainder of this paper is organized as follows.
In Section~\ref{sec:model_and_moments}, we introduce the model of an ABP under complete stochastic resetting in a harmonic trap. Using a Fokker-Planck-based approach, we derive the dynamical moment generating equation and apply the final renewal approach to calculate moments under stochastic resetting.
This equation is then used to compute the orientation autocorrelation and mean displacement in Section~\ref{sec:ncorr_ravg}.
In Section~\ref{sec:msd}, we determine the mean squared displacement and displacement fluctuation.
To examine deviations from Gaussian behavior over time, we calculate the fourth order moment of displacement and the excess kurtosis in Section~\ref{sec:excess_kurtosis}, and characterize the interplay of activity with resetting and trapping in the steady state phase diagrams in Section~\ref{sec:steady_state_phase_diagrams}.
Finally, we summarize the key results in Section~\ref{sec:conclusions}.

\section{Model and moments generator equation}
\label{sec:model_and_moments}
The standard active Brownian particle in two dimensions is described by its position $\mathbf{r} = (\mathrm{x}, \mathrm{y})$, and its orientation unit vector $\uv = (\mathrm{u}_x, \mathrm{u}_y)$ where $\mathrm{u}_x=\cos(\theta)$ and $\mathrm{u}_y=\sin(\theta)$, evolving over time $t$ from their initial values $(\rv_0,\uv_0)$. Stochastic resetting imposed on both $\rv$ and $\uv$ intermittently resets to initial values $(\rv_0,\uv_0)$ with rate $r$. The position $\rv$ and orientation $\uv$ evolves within a harmonic trap
\bea
&&\dot \rv = v_0 \uv + \sqrt{2D} \boldsymbol{\chi}(t)-\mu k \rv\,,
\label{eom:disp}\\
&&\dot \uv =  \sqrt{2D_r}\eta (t)\uv^{\perp}\,,
\label{eom:rot_active}\\
&&(\rv, \uv) \rightarrow (\rv_0, \uv_0)\,.
\label{eom:resetting}
\eea
Here, $\uv^{\perp}=(-\mathrm{u}_y, \mathrm{u}_x)$ is the normal unit vector to $\uv$, $v_0$ is the active speed, $D$ is the translation diffusion coefficient, $\mu$ is the mobility, and $k$ is the strength of the harmonic trap. $D_r$ is the orientation diffusion coefficient. The noise terms $\boldsymbol{\chi}(t)$ and $\eta(t)$ are modeled as Gaussian white noise with zero mean and variances given by $\la \chi_i(t) \chi_j(t^{\prime})\ra = \d_{ij} \d(t-t^{\prime})$ and $\la\eta(t)\eta(t^{\prime})\ra=\d(t-t^{\prime})$, respectively. The interplay of the three timescales $D_{r}^{-1}$, $r^{-1}$, and $(\mu k)^{-1}$ would lead to rich
behavior for ABP under resetting in a harmonic trap.

We rescale the dynamics using timescale $\t=D_r^{-1}$ and length scale $\ell = \sqrt{D/D_r}$. The dynamics ($\rv\to \rv/\ell$, $\uv\to\uv$, $t\to t/\t$) is now controlled by activity strength defined by P\'eclet $\Pe = v_0\t/\ell$, resetting rate $r=r\t$, and trapping strength $\beta=\mu k \ell$. We investigate the dynamics and steady state behavior with varying three dimensionless parameters activity $\Pe$, resetting rate $r$, and trap strength $\beta$.

\begin{figure}[!t]
\begin{center}
\includegraphics[width=7cm]{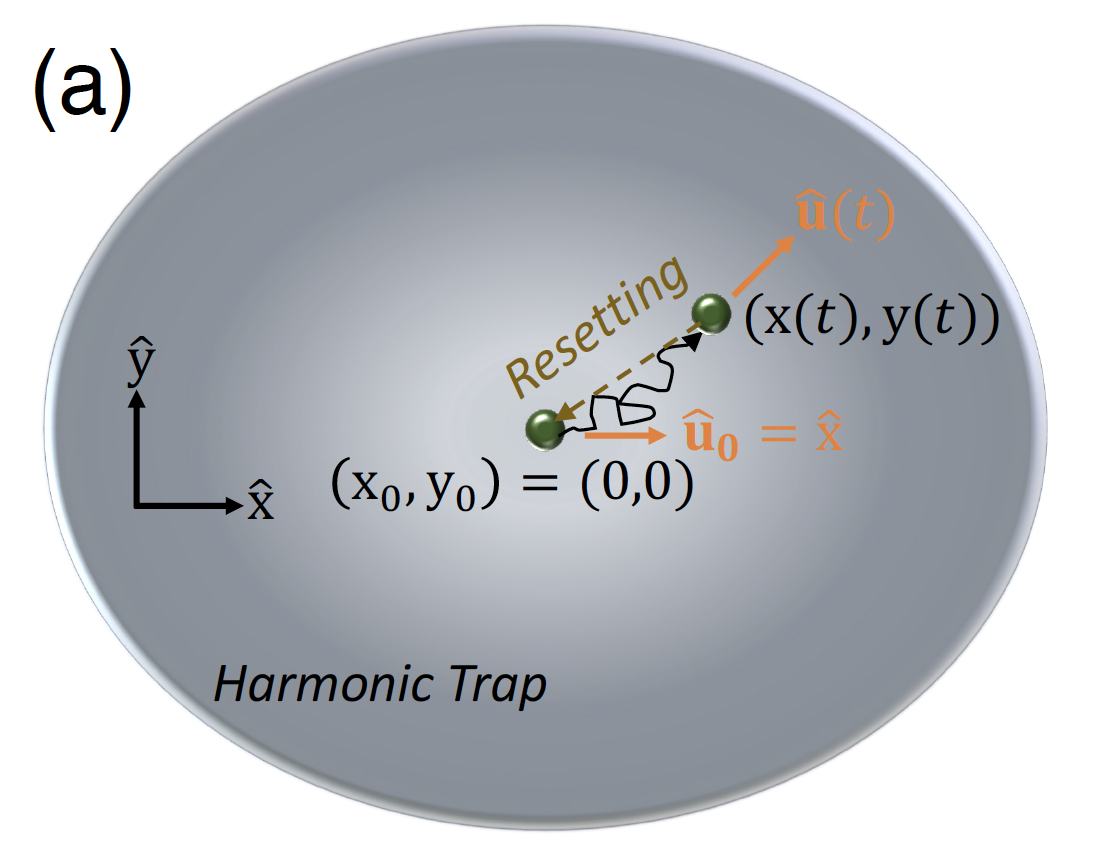}
\includegraphics[width=7cm]{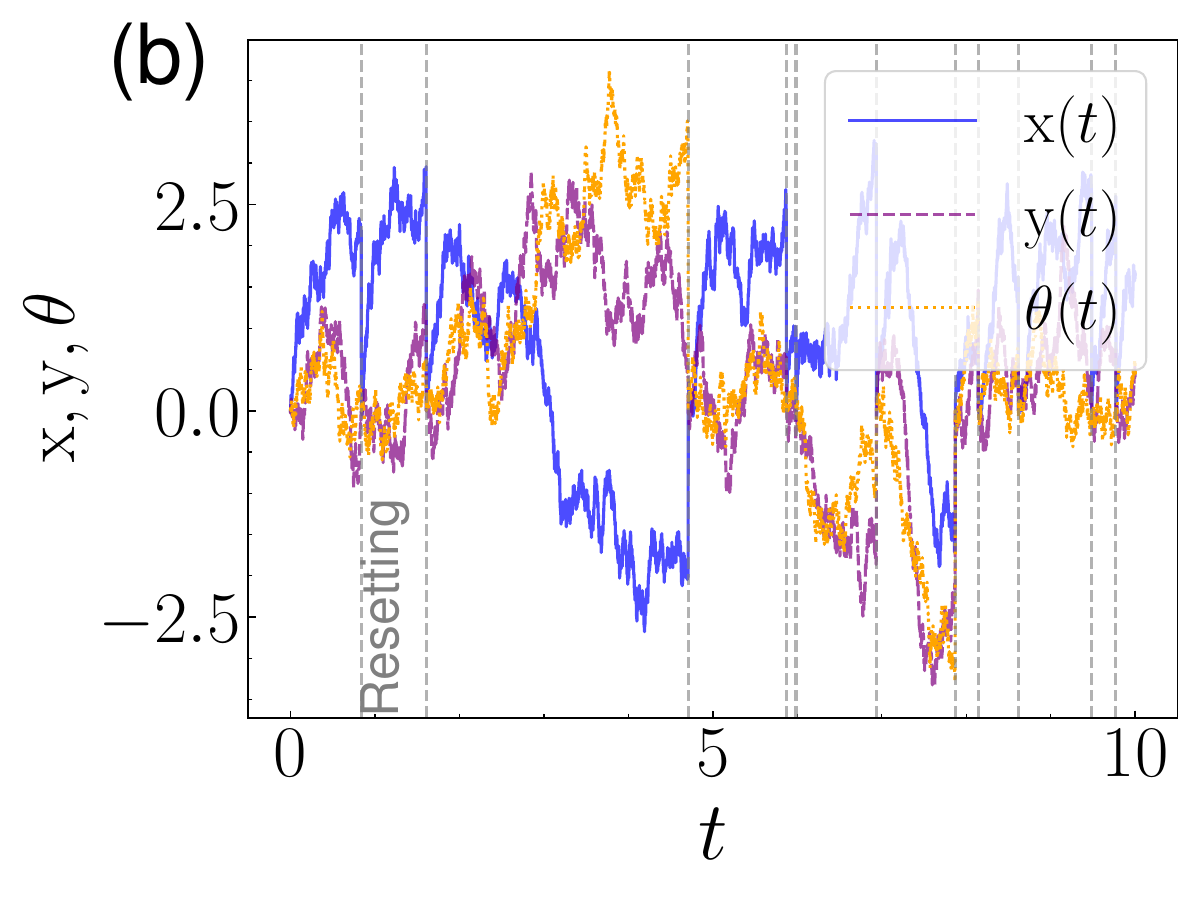}
\includegraphics[width=17cm]{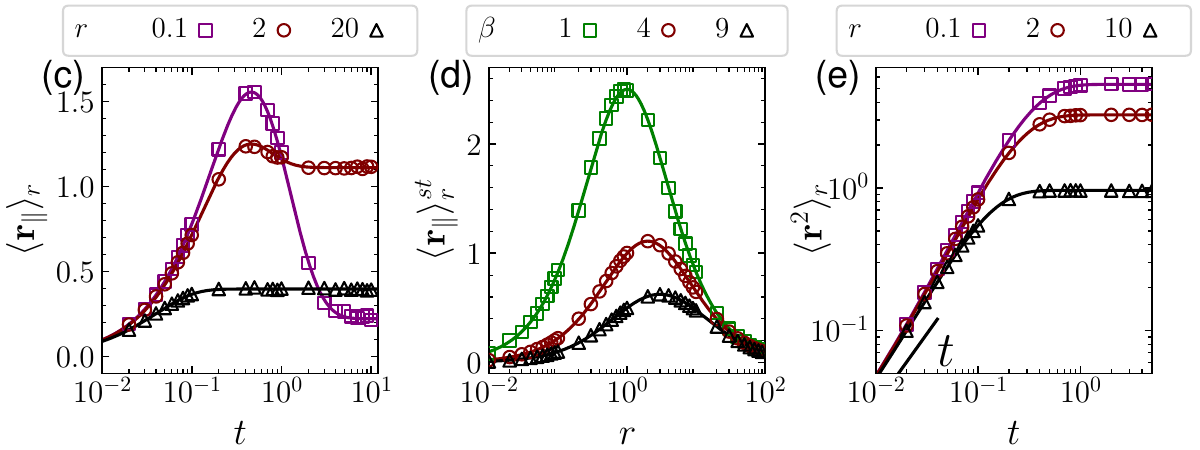} 
\caption{Schematic diagram, time series and and lower order moments. (a) Schematic illustration of an active Brownian particle under stochastic position and orientation resetting in a harmonic trap.
(b) Typical time evolution of $\mathrm{x}$, $\mathrm{y}$, and $\theta$ at $\Pe=10$, $\beta=4$, and $r=1$. The simulation time step used is $dt=0.001$. The vertical dashed lines indicate the random resetting events at which the particle’s state is reset to zero. The first reset event is labeled ``Resetting."
(c) Mean parallel displacement along initial orientation direction in two dimensions (equation~(\ref{eq:rpara_resetting})) for resetting rate $r=0.1 (\mathrm{square}, \Box),~2(\mathrm{circle}, \circ),~10(\mathrm{triangle},\triangle)$ with harmonic trap strength $\beta =4$ and activity $\Pe=10$.  
(d) Steady state displacement along initial orientation $\la\rv_{\parallel}\ra^{st}_r$ (equation~(\ref{eq:rpara_st})) as a function of $r$ for $\beta = 1 (\mathrm{square}, \Box), 4 (\mathrm{circle}, \circ), 9 (\mathrm{triangle},\triangle)$ with $\Pe=10$.
(e) Mean squared displacement (MSD) as function of time (equation~(\ref{eq:r2avg_resetting})) correspond to the parameters in (c).
The points represent the simulation results, while the lines correspond to the analytical solutions.
The initial and resetting position is at the origin with the initial and resetting orientation along the $\mathrm{x}$-axis.} 
\label{fig1}
\end{center}
\end{figure} 
\subsection*{Moments generator equation}
The probability distribution $P(\rv, \uv, t)$ of the position $\rv$ and the active orientation $\uv$ of the particle  follows the Fokker-Planck equation~\cite{Chaudhuri2021}
\bea
\p_t P(\rv, \uv, t) &=& \nabla^2 P +\nabla_\uv^2 P - \Pe\, \uv\cdot \nabla P + \beta \rv \cdot \nabla P + 2 \beta P\,,
\label{eq:F-P}
\eea
where $\nabla$ is the two dimensional Laplacian operator, and $\nabla_\uv$ is the Laplacian in the orientation space. 

Utilizing the Laplace transform $\tilde P(\rv, \uv, s) = \int_0^\infty dt\, e^{-s t}\, P(\rv, \uv, t) $ and defining the mean of an observable $\la \psi \ra_s = \int d\rv \, d\uv\, \psi(\rv, \uv ) \tilde P(\rv, \uv, s)$, multiplying by $\psi(\rv, \uv)$ and integrating over all possible $(\rv, \uv)$  we find~\cite{Chaudhuri2021},
\bea
-\la \psi \ra_0 + s \la \psi \ra_s &=& \la \nabla^2 \psi \ra_s + \la \nabla_\uv^2 \psi \ra_s + \Pe\la\, \uv\cdot \nabla \psi \ra_s - \beta \la \rv  \cdot \nabla \psi \ra_s\,,
\label{eq:moment}
\eea
where the initial condition sets $\la \psi \ra_0 = \int d\rv \, d\uv\,  \psi(\rv, \uv) P(\rv, \uv, 0)$. Without any loss of generality, we consider the initial condition to follow $P(\rv, \uv, 0) = \d(\rv-\rv_0) \d(\uv - \uv_0)$, where $\rv_0$ and $\uv_0$ are the initial position and orientation respectively. We use equation~(\ref{eq:moment}) to calculate exact moments as a function of time without resetting, which is already explored in Chaudhuri et al.~\cite{Chaudhuri2021}.

The moments of an ABP under stochastic position and orientation resetting in presence of a harmonic trap can be calculated exactly as~\cite{Kumar2020}
\bea
\la\psi(t)\ra_r &=& e^{-r t} \la\psi(t)\ra + r \int_{0}^{t} dt^{\prime} e^{-rt^{\prime}} \la\psi(t^{\prime})\ra\,.
\label{eq:moment_resetting}
\eea
We utilize equation~(\ref{eq:moment_resetting}) to compute the analytic moments exactly under resetting. For comparison with simulations, we perform Euler-Maruyama integration of equations~(\ref{eom:disp}) and (\ref{eom:rot_active}) with stochastic resetting equation~(\ref{eom:resetting}). The initial position of the particles set at the origin $(x,y) = (0,0)$ which corresponds to the minimum of the harmonic potential, with the orientation along the $x$-axis i.e., $\theta=0$. The positions and orientations of particle stochastically reset to their initial state $(x,y,\theta)=(0,0,0)$ at rate $r$.

\section{Mean orientation and mean displacement}
\label{sec:ncorr_ravg}
We consider the initial orientation of ABP along $\la\uv\ra=\uv_0$ and proceed to calculate mean orientation. To calculate $\la\uv\ra$, we use $\psi =\uv$ in the equation~(\ref{eq:moment}), leads to $\la\uv\ra_s = \uv_0/(1+s)$. Inverse Laplace transform leads to the average orientation without stochastic resetting $\la\uv(t)\ra = e^{-t} \uv_0$. In the presence of stochastic resetting using renewal approach in equation~(\ref{eq:moment_resetting}) and taking the dot product with the initial orientation, we get the orientation autocorrelation,
\bea
\la\uv(t)\cdot\uv(0)\ra_r &=& \f{r+e^{-(1+r)t}}{(1+r)}\,. 
\label{eq:ncorr_resetting}
\eea
The plot of equation~(\ref{eq:ncorr_resetting}) is compared with simulations, as shown in \ref{app:ncorr_rpara}, figure~\ref{app_fig1}(a).  
As ($t\to\infty$), the orientation autocorrelation saturates to $\la\uv(t)\cdot\uv(0)\ra^{st}_r=r/(1+r)$. In the steady state $\la\uv(t)\cdot\uv(0)\ra^{st}_r$ varies from $0$ as resetting rate $r\to 0$ to $1$ as $r\to\infty$.

Now, we proceed to calculate $\la\rv\ra_s$ we consider the initial position at $\la\rv\ra_0=0$, using the equation~(\ref{eq:moment}) leads to $\la\rv\ra_s = \Pe\la\uv\ra_s/(s+\beta)$. Substituting $\la\uv\ra_s$ and then inverse Laplace transform gives the mean displacement in the absence of resetting $\la\rv(t)\ra$~\cite{Chaudhuri2021}. The parallel and perpendicular component of the displacement vector to the initial orientation defined as $\uv_0$ as $\rv_{\parallel} = (\rv\cdot\uv_0)\uv_0$ and $\rv_{\perp} =\rv - \rv_{\parallel}$. In the presence of stochastic resetting using renewal approach in equation~(\ref{eq:moment_resetting}), we get
\bea
\fl\la\rv_{\parallel}(t)\ra_r  = \f{\Pe \uv_0 (e^{-(\beta+r) t} -e^{-(1+r)t})}{1-\beta} - \f{r\Pe\uv_0}{1-\beta} \left[\f{1-e^{-(1+r)t}}{1+r}-\f{1-e^{-(\beta+r)t}}{\beta+r}\right]\,.
\label{eq:rpara_resetting}
\eea
In the absence of harmonic trap ($\beta=0$), equation~(\ref{eq:rpara_resetting}) simplifies to ABP under stochastic resetting~\cite{Kumar2020}, see \ref{app:ncorr_rpara}, figure~\ref{app_fig1}(b).
In figure~\ref{fig1}(a), we plot equation~(\ref{eq:rpara_resetting}) as solid lines, showing excellent agreement with the simulation results represented by the points.
For a small resetting rate ($r$), $\la\rv_{\parallel}(t)\ra_r$ exhibits non-monotonic behavior: it starts with a small value in the short time, thermally dominated diffusion regime, reaches a maximum at intermediate times in the activity-dominated regime, and then decays back to smaller values due to trapping, see for $r=0.1$ (squares, $\Box$) in figure~\ref{fig1}(c). At long times,  $\la\rv_{\parallel}(t)\ra_r$  decays to zero in the absence of stochastic resetting~\cite{Chaudhuri2021}, exhibiting the same behavior as for low resetting rates.
For intermediate values of the resetting rate ($r$), $\la\rv_{\parallel}(t)\ra_r$ initially shows a small value in the short-time, thermally dominated diffusion regime, reaches a maximum at intermediate times in the activity-dominated regime, and then does not decay back to smaller values in long-time, see for $r=2$ (circles, $\circ$) in figure~\ref{fig1}(c).
For large values of the resetting rate ($r$), $\la\rv_{\parallel}(t)\ra_r$ initially shows a small value in the short-time, thermally dominated diffusion regime, slightly increases in the activity-dominated regime, and then saturates due to the dominance of stochastic resetting, see for $r=20$ (triangles, $\triangle$) in figure~\ref{fig1}(c).

At, long times $t\to\infty$, mean parallel displacement $\la\rv_{\parallel}\ra^{st}_r = \lim_{t\to\infty}\la\rv_{\parallel}(t)\ra_r$
\bea
\la\rv_{\parallel}\ra^{st}_r &=& \f{r\Pe\uv_0}{(r+\beta)(1+r)}\,.
\label{eq:rpara_st}
\eea
The mean parallel displacement at steady state show a local maximum at resetting rate $r_{\mathrm{max}}=\sqrt{\beta}$.
In figure~\ref{fig1}(d), we plot equation~(\ref{eq:rpara_st}) (solid lines) as a function of resetting rates ($r$), comparing it with simulation results (points). The maximum mean parallel displacement occurs at resetting rates  $r_{\mathrm{max}}=\sqrt{\beta}=1,~2,~3$ for trap strengths $\beta=1 ~(\mathrm{squares}, \Box),~4 ~(\mathrm{circles}, \circ),~9 ~(\mathrm{triangles}, \triangle)$, respectively. It is important to compute the second order moments to further quantify the fluctuating dynamics.

\begin{figure*}[!t]
\begin{center}
\includegraphics[width=17cm]{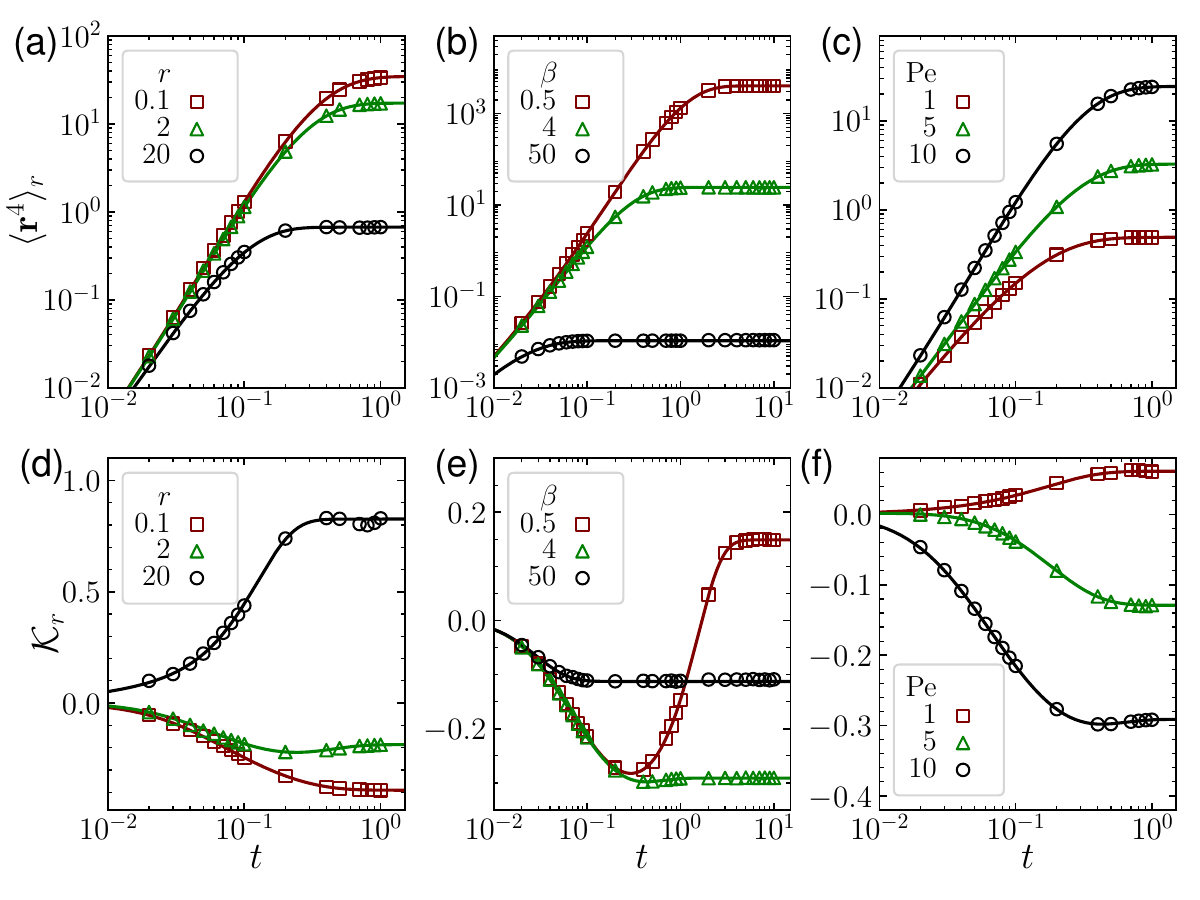} 
\caption{
Time evolution of the fourth order moment and excess kurtosis.
(a), (b), (c) Fourth order moment of displacement $\la\rv^4\ra_{r}$ (equation~(\ref{eq:r4avg_resetting})); (d), (e), (f) excess kurtosis $\mathcal{K}_r$ as a function of time $t$ for different conditions: (a),(d) resetting rates $r = 0.1~(\mathrm{square}, \Box), ~2~(\mathrm{triangle}, \triangle), ~20~(\mathrm{circle}, \circ)$ with activity $\Pe = 10$ and harmonic trap strength $\beta = 4$, (b),(e) $\beta = 0.5~(\mathrm{square}, \Box), ~4~(\mathrm{triangle}, \triangle), ~50~(\mathrm{circle}, \circ)$ with $\Pe = 10$ and $r = 1$, and (c),(f) $\Pe = 1~(\mathrm{square}, \Box), ~5~(\mathrm{triangle}, \triangle), ~10~(\mathrm{circle}, \circ)$ with $r = 1$ and $\beta = 4$.
The solid lines represent analytic predictions, while points denote simulation results.
The initial and resetting position is at the origin, with the initial and resetting orientation along the $x$-axis.
The region where $\mathcal{K}_r<0$ (red) represents the activity-dominated regime, and $\mathcal{K}_r>0$ (green) represents the resetting-dominated regime.
} 
\label{fig2}
\end{center}
\end{figure*} 

\section{Mean squared displacement (MSD)}
\label{sec:msd}

We proceed to compute mean squared displacement $\la\rv^2\ra_s$ defining observable $\psi=\rv^2$, utilizing equation~(\ref{eq:moment}) with initial position at origin $\la\rv^2\ra_0=0$ leads to the mean squared displacement in Laplace space $\la\rv^2\ra_s$ 
\bea
\la\rv^2\ra_s &=& \f{1}{s+2\beta}\left[\f{4}{s} + 2 \Pe \la\rv\cdot\uv\ra_s\right]~,
\label{eq:r2avg-Laplace}
\eea
where the second term $\la\rv\cdot\uv\ra_s$ calculated assuming $\psi=\rv\cdot\uv$ in equation~(\ref{eq:moment}) as
\bea
\la\rv\cdot\uv\ra_s=\f{\Pe}{s(s+\beta+1)}~.
\eea 
Inverse Laplace transform of equation~(\ref{eq:r2avg-Laplace}) gives mean squared displacement $\la\rv^2(t)\ra$ without stochastic resetting (see~\ref{app:MSD}), which previously obtained in Chaudhuri et al.~\cite{Chaudhuri2021}. Finally, using renewal approach in equation~(\ref{eq:moment_resetting}), we get 
\bea
\fl\la\rv^2(t)\ra_r = e^{-r t} \left[\f{2}{\beta}\left(1-e^{-2\beta t}\right) +\f{\Pe^2}{\beta(1+\beta)} - \f{2\Pe^2 e^{-\beta t}}{1-\beta} \left[\f{e^{-\beta t}}{2\beta} - \f{e^{-t}}{1+\beta}\right]\right] \nn\\
\fl+ \f{(2+2\beta+\Pe^2)(1-e^{-rt})}{\beta (1+\beta)} + \f{2r\beta\Pe^2[1-e^{-(1+\beta+r)t}]}{(1+\beta+r)(1-\beta^2)} + \f{r(\Pe^2+2\beta-2)(1-e^{-(2\beta+r)t})}{\beta(2\beta+r)(1-\beta)}\,.\nonumber\\
\label{eq:r2avg_resetting}
\eea
In figure~\ref{fig1}(e), we compare the equation~(\ref{eq:r2avg_resetting}) represented by solid lines, with simulations depicted by points, showing excellent agreement. The various limiting cases of equation~(\ref{eq:r2avg_resetting}) depending on parameters $r$, $\Pe$, and $\beta$ discussed in \ref{app:MSD}.

\begin{figure*}[!t]
\begin{center}
\includegraphics[width=17cm]{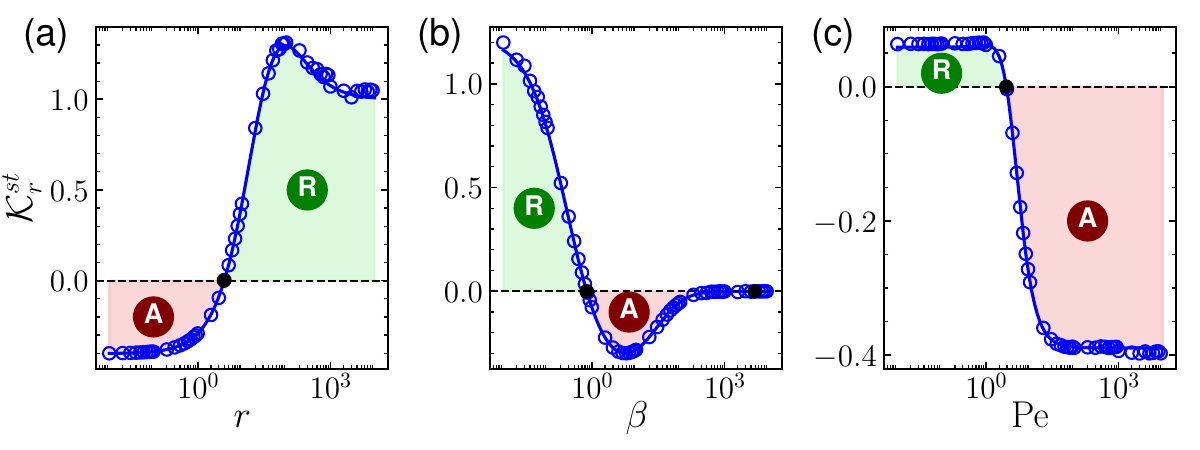} 
\includegraphics[width=17cm]{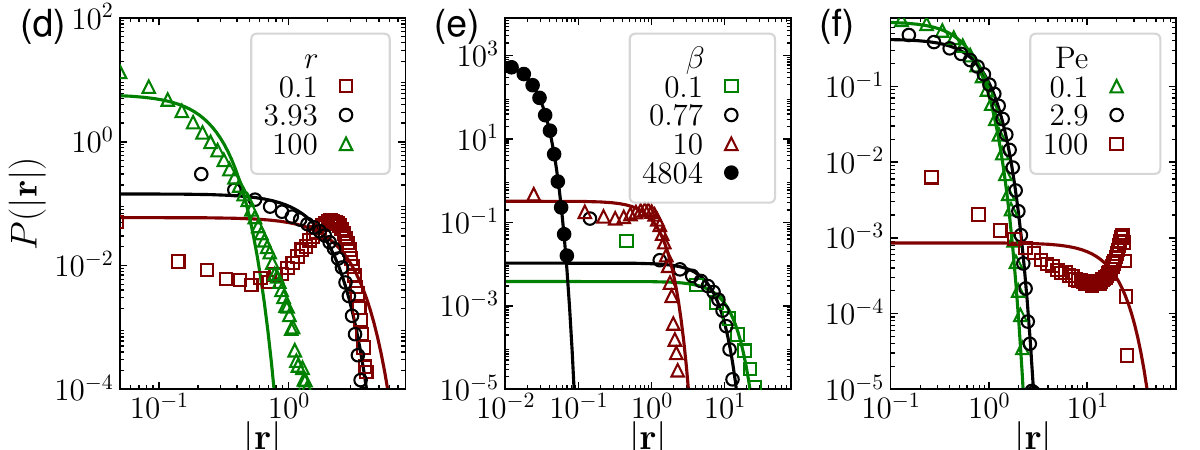} 
\caption{
Steady state excess kurtosis, $\mathcal{K}^{st}_{r}$, as a function of (a) resetting rate $r$ with activity $\Pe=10$ and harmonic trap strength $\beta=4$, (b) $\beta$ with $\Pe=10$ and $r=1$, and (c) $\Pe$ with $r=1$ and $\beta=4$.
The blue solid line represents the exact analytical solution for steady state excess kurtosis as given by equation~(\ref{eq:kurtosis_st}).
The symbols (blue open circles) are from simulation results.
The regions where $\mathcal{K}^{st}_{r}<0$ (red) correspond to the activity (A)-dominated regime, while regions where $\mathcal{K}^{st}_{r}>0$ (green) correspond to the resetting (R)-dominated regime.
The critical points where $\mathcal{K}^{st}_{r}=0$ (black solid symbols) occur at $r_c=3.93$ in (a), $\beta^{I}_c=0.77$ and $\beta^{II}_c=4804$ in (b), and $\Pe_c=2.90$ in (c).
The radial probability distributions for (d) three resetting rates ($r$) correspond to (a), (e) four harmonic trap strengths ($\beta$) to (b), and (f) three activity values ($\Pe$) to (c).
(d, e, f) The points represent simulation results, and the solid lines depict the Gaussian probability distribution derived from equation~(\ref{eq:r2avgr_st}).} 
\label{fig3}
\end{center}
\end{figure*}

\noindent
At small time ($t\to 0$), equation~(\ref{eq:r2avg_resetting}) leading to
\bea
\fl\la\rv^2\ra_r \simeq 4t + (\Pe^2 -4\beta -2 r) t^2 + \f{1}{3}\left[8\beta^2 - \Pe^2 - 3 \beta \Pe^2 + 8\beta r -2\Pe^2 r +2r^2\right]t^3+\mathcal{O}(t^4)\,.\nonumber\\
\label{eq:r2avg_resetting_small_times}
\eea
The MSD at small time exhibits diffusive behavior $\la\rv^2\ra_r=4t$, which crosses over to ballistic behavior $\la\rv^2\ra_r\sim t^2$ when $\Pe^2>2r+4\beta$ at $t^{*}=4/(\Pe^2-2r-4\beta)$. 
In the steady state, MSD $\la\rv^2\ra^{st}_r = \la\rv^2\ra_r(t\to\infty)$ gives
\bea
\la\rv^2\ra^{st}_r &=& \f{4}{2\beta+r}+\f{2\Pe^2}{(1+\beta+r)(2\beta+r)}~.
\label{eq:r2avgr_st}
\eea
The standard deviation is given by $\sigma=\sqrt{\la\rv^2\ra^{st}_r/2}$, and the Gaussian probability distribution is $P(|\rv|) = e^{-\rv^2/2\sigma^2} / 2\pi \sigma^2$. The latter is represented by solid lines in figures~\ref{fig3}(d), \ref{fig3}(e), and \ref{fig3}(f). We use this Gaussian distribution to demonstrate deviations in the presence of non-equilibrium parameters $r$ and $\Pe$.

\noindent
The effective diffusion coefficient $D_{\rm eff} = (\beta+r) \la \rv^2 \ra_{\rm st}/2$
\bea
D_{\rm eff} &=& \f{2(\beta+r)}{2\beta+r}+\f{(\beta+r)\Pe^2}{(1+\beta+r)(2\beta+r)}\,.
\eea
Setting resetting $r=0$ and activity $\Pe=0$, $D_{\rm eff}=1$ holds the fluctuation dissipation relation (FDR).
Excess diffusion coefficient due to non-equilibrium nature, calculated as $D_{\rm ex} = D_{\rm eff} - 1$ reads
\bea
D_{\rm ex} &=& \f{r}{2\beta+r}+\f{(\beta+r)\Pe^2}{(1+\beta+r)(2\beta+r)}\,.
\eea
The fluctuation-dissipation relation holds for $D_{\rm ex}=0$, i.e., for $r=\Pe=0$. Both activity ($\Pe$) and resetting ($r$) break this relation and result in $D_{\rm ex}>0$. However, it is not possible to precisely distinguish between the activity-dominated and resetting-dominated regimes by calculating $D_{\rm ex}$ when varying the parameters $r, \beta, \Pe$. We distinguish them by calculating the excess kurtosis in the next section.

\noindent
In the absence of stochastic resetting ($r=0$), we get ABP in a harmonic trap~\cite{Chaudhuri2021}. In the absence of activity ($\Pe=0$), Brownian particle in two dimensions under resetting gives $\la\rv^2\ra^{st}_r=4/(2\beta+r)$, $D_{\rm eff}=2(\beta+r)/(2\beta+r)$, and $D_{\rm ex}=r/(2\beta+r)$.

Moreover, the displacement fluctuations and their limiting cases are discussed in \ref{app:disp_fluct}. The components of the second order moment, such as those along the initial orientation direction, are also discussed in \ref{app:second_moment_parallel}.

\section{Excess kurtosis: signature of non-Gaussian behavior}
\label{sec:excess_kurtosis}
To calculate the excess kurtosis, we first need to compute the fourth order moment of displacement.
%
Using a similar approach as for the lower order moments, we calculate the fourth order moment of displacement under stochastic position and orientation resetting. The detailed derivation is provided in \ref{app1} and final result for $\la\rv^4(t)\ra_r$ is shown in equation~(\ref{eq:r4avg_resetting}).
In the small time expansion ($t\to 0$), the fourth moment of displacement $\la\rv^4\ra_r=\lim_{t\to 0} \la\rv^4(t)\ra_r$ is given by
\bea
\fl\la\rv^4\ra_r \simeq 32 t^2 +\f{16}{3} \left(3\Pe^2-4r-12\beta\right)t^3 \nonumber\\
\fl+ \f{1}{3}\left[224 \beta^2+3 \Pe^4+24 r^2 +144\beta r -96 \beta \Pe^2 -16 \Pe^2-36r \Pe^2\right]t^4+\mathcal{O}(t^5)\,,
\eea
exhibits small time diffusive behavior $\la\rv^4\ra_r=32t^2$, which crosses over to ballistic behavior $\la\rv^4\ra_r\sim t^3$ at $t^{*}=6/(3\Pe^2-4r-12\beta)$ with $\Pe^2>4(r+3\beta)/3$. This crossover timescale, $t^{*}$, is smaller than the diffusive-to-ballistic crossover timescale calculated using the MSD in equation~(\ref{eq:r2avg_resetting_small_times}). This suggests that the MSD, exhibiting $\la\rv^2\ra_r\sim t$ to $\la\rv^2\ra_r\sim t^2$ crossovers, occurs later in time than $\la\rv^4\ra_r$, which shows $\la\rv^4\ra_r\sim t^2$ to $\la\rv^4\ra_r\sim t^3$ crossovers, rather than following $\la\rv^4\ra_r\sim (\la\rv^2\ra_r)^2$ in ballistic regime, which would exhibit $\la\rv^4\ra_r\sim t^2$ to $\la\rv^4\ra_r \sim t^4$ crossovers. It also indicates the onset of non-Gaussian effects, which can be more precisely predicted using the excess kurtosis involving both second-order moment and fourth-order moment of displacement, discussed later in the section.

In figures \ref{fig2}(a), \ref{fig2}(b), and \ref{fig2}(c), we have shown the $\la\rv^4\ra_r$ as a function of time $t$ for resetting rates $r=0.1,~2,~20$ with $\Pe=10$ and $\beta=4$ in (a), for trap strengths $\beta=0.5,~4,~50$ with $\Pe=10$ and $r=1$ in (b), and for activity values $\Pe=1,~5,~10$ with $\beta=4$ and $r=1$ in (c). The solid lines are the plot of equation~(\ref{eq:r4avg_resetting}) which excellently agrees with the simulation results depicted by points. 
In figures \ref{fig2}(a) and \ref{fig2}(b), we can see the increase of resetting rates $r$ and trapping strengths $\beta$ decreases the $\la\rv^4\ra_r$. On the other hand, the increase of activity parameter $\Pe$ increases the $\la\rv^4\ra_r$.
In the long time limit ($t\to\infty$), $\la\rv^4\ra_r$ reaches a steady state due to the effect of the harmonic trap and/or stochastic resetting. The steady state fourth moment of displacement $\la\rv^4\ra^{st}_r$ 
\bea
\fl\la\rv^4\ra^{st}_r = \frac{8}
{(1 + \beta + r) (2 \beta + r) (4 + 2 \beta + r) (1 + 3 \beta + r) (4 \beta + r)}\times\nonumber\\
\fl\left[48 \beta^3 + 8 (1 + r)^2 (4 + r) + \Pe^4 (8 + 3 r) + 8 \beta^2 (20 + 6 \Pe^2 + 11 r) \right.\nonumber\\
\fl \left.+ 4 \Pe^2 (8 + 14 r + 3 r^2) + 
     2 \beta \left[3 \Pe^4 + 8 \Pe^2 (7 + 3 r) + 24 (3 + 4 r + r^2)\right]\right]\,.   
\eea

To quantify the impact of thee control parameters activity ($\Pe$), resetting rate ($r$), and harmonic trap strength ($\beta$) on position distributions over time and at steady state, we calculate the excess kurtosis, which measures the deviation from a Gaussian process.
%
In the absence of activity ($\Pe=0$) and stochastic resetting ($r=0$), Brownian particle in two dimensions in a harmonic trap yields zero excess kurtosis, indicating Gaussian process.  We calculate excess kurtosis in the presence of activity and stochastic resetting, which deviates from zero.
Thus, the non-equilibrium nature of the system leads to a deviation from Gaussian behavior, as indicated by the excess kurtosis~\cite{Shee2020, Chaudhuri2021},
\bea
\mathcal{K}_r &=& \f{\la\rv^4\ra_r}{2\la\rv^2\ra_{r}^2} - 1\,.
\label{eq:kurtosis}
\eea
The fourth order moment of displacement $\la\rv^4\ra_r$ and mean squared displacement $\la\rv^2\ra_r$ have already been calculated as shown in equation~(\ref{eq:r4avg_resetting}) and equation~(\ref{eq:r2avg_resetting}), respectively. We expand the excess kurtosis in the small time limit ($t\to 0$),
\bea
\mathcal{K}_r &\simeq& \f{rt}{3} - \f{(3\Pe^4+16\beta r-4r \Pe^2)}{96} t^2 + \mathcal{O}(t^3)~.
\label{eq:kurtosis_t0}
\eea
The next order term $\mathcal{O}(t^3)$ is shown in equation~(\ref{eq:kurtosis_small_time}). It has shown the deviation of kurtosis towards positive or negative values in the small time controlled by $r$, $\Pe$, and $\beta$. The positive deviation holds for small activity ($\Pe\to 0$) but it deviates towards negative values with increase of activity.
For high activity, the excess kurtosis deviates towards negative values at $t^{*}=32r/(3\Pe^4+16\beta r -4 r \Pe^2)$.

In figures \ref{fig2}(d), \ref{fig2}(e), and \ref{fig2}(f), we have shown the $\mathcal{K}_r$ as a function of time $t$ for resetting rates $r=0.1,~2,~20$ with $\Pe=10$ and $\beta=4$ in (d), for trapping strength $\beta=0.5,~4,~50$ with $\Pe=10$ and $r=1$ in (e), and for P\'eclet values $\Pe=1,~5,~10$ with $\beta=4$ and $r=1$ in (f). The solid lines are the plot of equation~(\ref{eq:kurtosis}) which excellently agrees with the simulation results depicted by points. 
We characterize the activity- and resetting-dominated time regimes using excess kurtosis ($\mathcal{K}_r$). At short times, thermal diffusion leads to near-Gaussian behavior ($\mathcal{K}_r \sim 0$). Negative $\mathcal{K}_r$ indicates a flat, light-tailed with non-zero peak at $\Pe/\beta$ position distribution dominated by activity ($\Pe$), while positive $\mathcal{K}_r$ suggests a heavy-tailed with peak at zero position distribution dominated by resetting ($r$).
In figure \ref{fig2}(d), we clearly observe activity-dominated behavior ($\mathcal{K}_r < 0$) at low resetting rates ($r$) and resetting-dominated behavior ($\mathcal{K}_r > 0$) at high resetting rates ($r$).
In figure \ref{fig2}(e), we observe non-monotonic behavior at small trap strength ($\beta$). At short times, $\mathcal{K}_r$ shifts to negative values, indicating activity-dominated behavior, crosses zero at intermediate times, and eventually saturates to positive values, reflecting long-term resetting-dominated behavior.
The transitions from small to long times are: Gaussian ($\mathcal{K}_r=0$) to light-tail ($\mathcal{K}_r<0$), back to Gaussian ($\mathcal{K}_r=0$), and finally to heavy-tail ($\mathcal{K}_r>0$) position distributions.
In figure \ref{fig2}(f), we observe activity-dominated behavior ($\mathcal{K}_r < 0$) at high activity ($\Pe$) and resetting-dominated behavior ($\mathcal{K}_r > 0$) at low activity ($\Pe$).

To quantify the steady state properties of position distributions and capture the transitions between activity- and resetting-dominated regions, we calculate the steady state excess kurtosis and analyze the numerically obtained position distributions.

\section{Steady state excess kurtosis and phase diagrams}
\label{sec:steady_state_phase_diagrams}

\begin{figure*}[!t]
\begin{center}
\includegraphics[width=17cm]{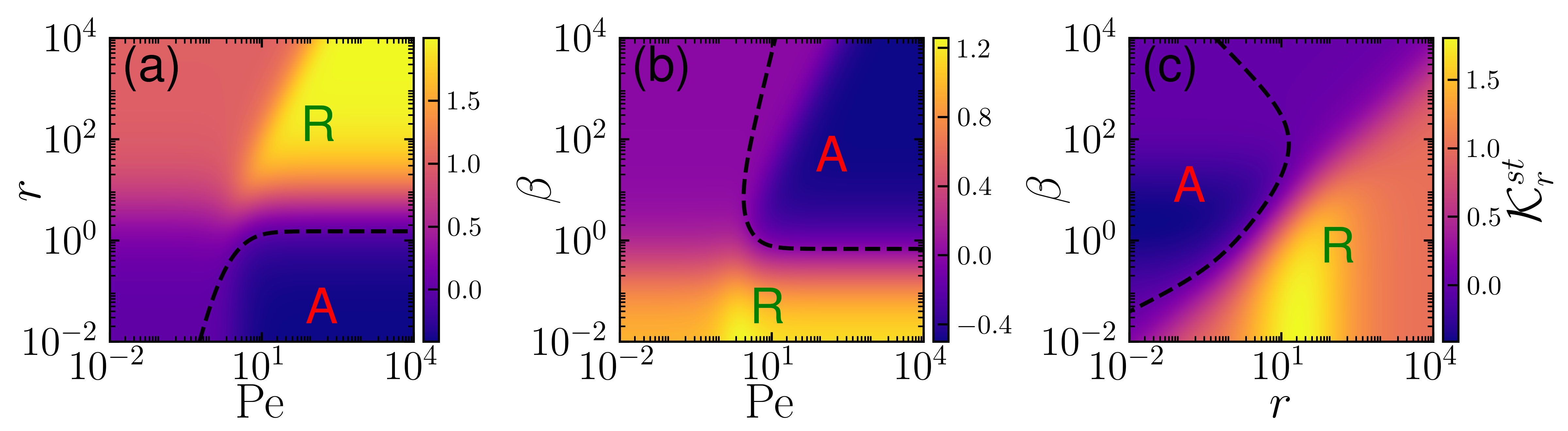} 
\caption{
Phase Diagrams.
Steady state excess kurtosis, $\mathcal{K}^{st}_{r}$, is shown in (a) activity($\Pe$)-resetting rate($r$) plane with harmonic trap strength $\beta=1$, (b) $\Pe-\beta$ plane with $r=1$, and (c) $r-\beta$ plane with $\Pe=10$.
The black dashed line corresponds to $\mathcal{K}^{st}_{r}=0$.
The region where $\mathcal{K}^{st}_{r}<0$ is denoted as the activity-dominated regime (A), and the region where $\mathcal{K}^{st}_{r}>0$ is denoted as the resetting-dominated regime (R).
} 
\label{fig4}
\end{center}
\end{figure*}

At long times($t\to\infty$), equation~(\ref{eq:kurtosis}) results steady state excess kurtosis $\mathcal{K}^{st}_r=\lim_{t\to\infty} \mathcal{K}_r(t)$
\bea
\fl\mathcal{K}^{st}_r = \f{2}{(4+2\beta+r)(1+3\beta+r)(4\beta+r)(2+2\beta+\Pe^2+2r)^2}\times\nonumber\\
\fl\left[12\beta^4 r -\beta^3 (6\Pe^4-20\Pe^2 r-2r(26+17r))\right.\nn\\
\fl\left.+r(1+r)(\Pe^4(2+r)+2(1+r)^2(4+r)+2\Pe^2(2r^2+9r+4))\right.\nn\\
\fl\left.+\beta^2(-\Pe^4(14+r)+2\Pe^2 r(32+19r)+2r(17r^2+55r+38))\right.\nn\\
\fl\left.+\beta r (\Pe^4(1+3r)+2(1+r^2)(22+7r)+\Pe^2(22r^2+80r+52))\right]\,.
\label{eq:kurtosis_st}
\eea
We now calculate various limiting cases to explore the properties of excess kurtosis.
\begin{enumerate}
\item  Brownian particle in a harmonic trap :
In the absence of activity ($\Pe=0$) and stochastic resetting ($r=0$), Brownian particle in two dimensions in a harmonic trap gives $\mathcal{K}_{r}^{st}=0$. It signifies the Gaussian process, and we calculate the steady state excess kurtosis to quantify the deviation from $\mathcal{K}_{r}^{st}=0$.

\item  Brownian particle under stochastic resetting : 
In the absence of activity ($\Pe=0$) and harmonic trap ($\beta=0$), Brownian particle in two dimensions under stochastic position and orientation resetting gives $\mathcal{K}^{st}_r=1$. The positive excess kurtosis signifies the heavy tailed position distribution at long times.

\item Active Brownian particle (ABP) :
In the absence of stochastic resetting ($r=0$) and harmonic trap ($\beta=0$), active Brownian particle (ABP) gives long time diffusive behavior with $\mathcal{K}^{st}_r=0$~\cite{Shee2020}.

\item Brownian particle under stochastic resetting in a harmonic trap :
In the absence of activity ($\Pe=0$), Brownian particle in two dimensions under stochastic position and orientation resetting in a harmonic trap gives $\mathcal{K}^{st}_r=r/(4\beta+r)$. It suggests the decrease of excess kurtosis with the increase of trapping strength $\beta$. For very high resetting rates, $\mathcal{K}^{st}_r$ reaches a maximum value of $1$.

\item ABP under stochastic resetting :
In the absence of harmonic trap ($\beta=0$), ABP under complete stochastic resetting, equation~(\ref{eq:kurtosis_st}) gives,
\bea
\fl\mathcal{K}^{st}_r = \f{\left[4(1+r)^2(4+r)+4(2r^2+9r+4)\Pe^2+2(2+r)\Pe^4\right]}{(4+r)(2(1+r)+\Pe^2)^2}~.
\label{eq:kurtosis_st_wo_ht}
\eea
Remarkably, in the absence of a harmonic trap ($\beta=0$), equation~(\ref{eq:kurtosis_st_wo_ht}) yields only positive values, indicating heavy-tailed position distributions (see \ref{app:abps_stochastic_resetting_wo_ht}).  

\item ABP in a harmonic trap :
In the absence of stochastic resetting ($r=0$), ABP in a harmonic trap gives the steady state excess kurtosis $\mathcal{K}_{r}^{st}$ already explored in~\cite{Chaudhuri2021}.
\end{enumerate}

In figures \ref{fig3}(a), \ref{fig3}(b), and \ref{fig3}(c), the steady state excess kurtosis $\mathcal{K}^{st}_r$ from equation~(\ref{eq:kurtosis_st}) is shown as blue solid lines, compared with simulation results represented by blue open points. The black solid points denote the Gaussian behavior at $\mathcal{K}^{st}_r=0$.
The negative excess kurtosis $\mathcal{K}^{st}_r < 0$ indicates persistent active motion, leading to a flat, light-tailed position distribution with a non-zero peak. This is referred to as the activity (A) dominated regime.
Positive excess kurtosis, $\mathcal{K}^{st}_r > 0$, indicates passive motion, resulting in a narrow, heavy-tailed position distribution peaking at zero. This is referred to as the resetting (R) dominated regime.
In figures \ref{fig3}(d), \ref{fig3}(e), and \ref{fig3}(f), we plot the radial distribution from simulations (points) alongside the corresponding Gaussian distribution (solid lines) using the second order moment.

Figure \ref{fig3}(a) shows a monotonic transition from the activity (A) dominated regime at small resetting rates ($r$) to the resetting (R) dominated regime at larger resetting rates, with a critical resetting rate ($r_c$) marking Gaussian behavior. Interestingly, in the resetting dominated regime, the steady state excess kurtosis $\mathcal{K}^{st}_r$ exhibits a non-monotonic transition, reaching its maximum positive value at intermediate $r$. This is purely due to the interplay between activity ($\Pe$) and resetting ($r$), which occurs even in the absence of a harmonic trap (see \ref{app:abps_stochastic_resetting_wo_ht}). In figure \ref{fig3}(d), we present the radial distribution for three resetting rates: the activity-dominated regime at $r < r_c$, Gaussian behavior at $r = r_c$, and the resetting-dominated regime at $r > r_c$. We clearly observe deviation in the radial distribution compared to the solid lines representing the Gaussian probability distribution derived from the corresponding steady-state mean-squared displacement in equation~(\ref{eq:r2avgr_st}). The radial distribution, represented by points from simulations, exhibits a non-zero peak with a light-tailed profile (negative excess kurtosis, activity-dominated regime) for small resetting rates ($r<r_c$) and a peak at zero with a heavy-tailed profile (positive excess kurtosis, resetting-dominated regime) for large resetting rates ($r>r_c$).

Figure \ref{fig3}(b) shows a non-monotonic transition as a function of trap strength $\beta$. Initially, the resetting (R) dominated regime occurs at weak trap strength, transitioning to the activity (A) dominated regime at stronger trap strengths, with a critical trap strength ($\beta^{I}_c$) indicating Gaussian behavior. Subsequently, the activity (A) dominated regime transitions back to the resetting regime at very strong trap strengths, marked by another critical trap strength ($\beta^{II}_c$). In figure \ref{fig3}(e), we present the radial distribution for four trap strengths: the resetting-dominated regime at $\beta < \beta_c$, Gaussian behavior at $\beta = \beta^{I}_c$, the activity-dominated regime at $\beta > \beta_c$, Gaussian behavior at $\beta = \beta^{II}_c$. In the resetting-dominated regime for $\beta > \beta^{II}_c$, the steady state excess kurtosis $\mathcal{K}^{st}_r$ takes very small positive values, making the heavy-tail visualization nearly impossible.

Figure \ref{fig3}(c) shows a monotonic transition from the resetting (R) dominated regime at small activity ($\Pe$) to the activity (A) dominated regime at larger activity, with a critical resetting rate ($\Pe_c$) marking Gaussian behavior. In figure \ref{fig3}(f), we present the radial distribution for three activity ($\Pe$) values: the resetting-dominated regime at $\Pe < \Pe_c$, Gaussian behavior at $\Pe = \Pe_c$, and the activity-dominated regime at $\Pe > \Pe_c$.

In figures \ref{fig4}(a), \ref{fig4}(b), and \ref{fig4}(c), we plot the phase diagram considering $\mathcal{K}^{st}_r$ as order parameter to show the activity (A) and resetting (R) dominated regime separating by Gaussian line $\mathcal{K}^{st}_r=0$ depicted by black dashed line. 
Figure \ref{fig4}(a) shows the phase diagram in the activity($\Pe$)-resetting rate($r$) plane, highlighting two key points: at small resetting rates ($r$), the transition progresses from weak resetting (R) to Gaussian, then to the activity (A) dominated regime. At large resetting rates, increasing activity enhances the resetting (R) dominated regime, resulting in a heavier tail in the distribution. 
Figure \ref{fig4}(b) shows the phase diagram in the activity($\Pe$)-harmonic trap strength($\beta$) plane, highlighting a re-entrant transition with increasing trap strength ($\beta$), progressing from resetting (R) to Gaussian, then to activity (A), back to Gaussian, and finally returning to resetting-dominated behavior. The suppression of resetting behavior with increasing trap strength in figure \ref{fig4}(b) is also illustrated in the resetting rate($r$)-harmonic trap strength($\beta$) plane in figure \ref{fig4}(c). The resetting (R)-dominated regime can be classified into two subcategories based on effective particle behavior: 1) Brownian particles under stochastic resetting in a harmonic trap, with steady-state excess kurtosis $0 < \mathcal{K}^{st}_r < 1$, and 2) Active Brownian particles (ABP) under stochastic resetting, where $\mathcal{K}^{st}_r > 1$. The line $\mathcal{K}^{st}_r = 1$ marks the critical transition between these two regimes, where the particle behaves as a Brownian particle under stochastic resetting.

\section{Conclusions}
\label{sec:conclusions}

In this work, we computed the exact analytical moments of a two dimensional Active Brownian Particle (ABP) subject to complete stochastic resetting (both position and orientation) in a harmonic trap. These analytical results were validated against numerical simulations, demonstrating excellent agreement, and the nonequilibrium steady state behavior was thoroughly analyzed.

In the steady state, the orientation autocorrelation decays to a non-zero value in the presence of resetting, increasing with the resetting rate and approaching unity as the resetting rate becomes large. The steady state mean displacement exhibits non-monotonic behavior with resetting rate, initially increasing from zero as the resetting rate rises, peaking, and then returning to near-zero at very high resetting rates. We showed that ballistic dynamics emerge in the mean squared displacement (MSD) at intermediate times with increasing activity, though these are suppressed when resetting dominates. Additionally, the MSD reveals that the fluctuation-dissipation relation (FDR) holds in the absence of both activity and resetting, but breaks down, yielding a non-zero excess diffusion coefficient, when either is present.

To distinguish between the activity-dominated (A) and resetting-dominated (R) regimes, we computed the fourth moment of displacement and the corresponding excess kurtosis, providing a complete characterization of the system. In the steady state, excess kurtosis captures the transition between the activity-dominated regime (negative excess kurtosis) and the resetting-dominated regime (positive excess kurtosis), with the Gaussian regime represented by zero excess kurtosis. We anticipate that this rich steady-state behavior can be experimentally validated in active colloids or robotic systems, as shown in~\cite{Paramanick2024}, and could also serve as a basis for understanding the homing strategies of living organisms, from animals to birds. In the future, it will be crucial to explore active Brownian particle (ABP) under two additional stochastic resetting protocols~\cite{Kumar2020, Baouche2024}, only position resetting and only orientation resetting in a harmonic trap, to gain valuable insights into optimal resetting strategies in external potentials.

\section*{Acknowledgements}
AS acknowledges partial financial support from the John Templeton Foundation, Grant 62213.

\appendix

\section{Orientation autocorrelation and mean displacement without harmonic trap}
\label{app:ncorr_rpara}
\begin{figure}[!t]
\begin{center}
\includegraphics[width=17cm]{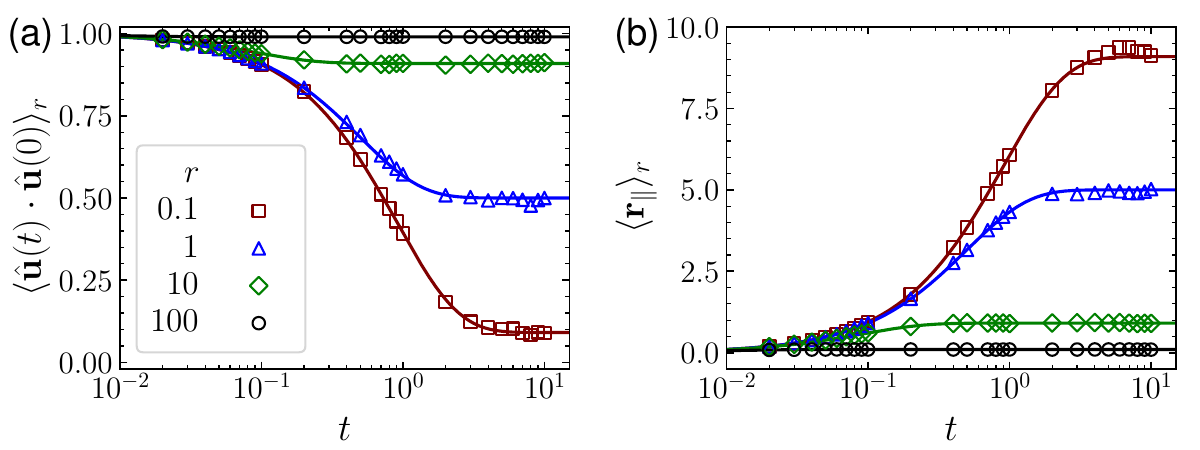} 
\caption{
(a) Orientation autocorrelation (equation~(\ref{eq:ncorr_resetting})) and (b) mean parallel displacement (equation~(\ref{eq:rpara_resetting})) as a function of time in two dimensions for $r=0.1 (\mathrm{square}, \Box),~1(\mathrm{triangle},\triangle),~10(\mathrm{diamond},\Diamond),~100(\mathrm{circle}, \circ)$ with $\beta =0$ and $\Pe=10$.  
Points are from simulation results and solid lines are analytic plot.
The initial and resetting position is at the origin with the initial and resetting orientation along the $x$-axis.
} 
\label{app_fig1}
\end{center}
\end{figure} 
In figure~\ref{app_fig1}(a), we plot the orientation autocorrelation from equation~(\ref{eq:ncorr_resetting}) for varying resetting rates ($r$). As the resetting rate increases, the orientation autocorrelation saturates at higher, non-zero values.
In figure~\ref{app_fig1}(b), we plot the mean parallel displacement as a function of time from equation~(\ref{eq:rpara_resetting}) for varying resetting rates ($r$) with activity $\Pe=10$ in the absence of a harmonic trap ($\beta=0$).

\section{Limiting cases of mean squared displacement (MSD)}
\label{app:MSD}
\subsection{ABP in a harmonic trap}

Inverse Laplace transform of equation~(\ref{eq:r2avg-Laplace}) gives mean squared displacement $\la\rv^2(t)\ra$ for an ABP in a harmonic trap without stochastic resetting ($r=0$ in equation~(\ref{eq:r2avg_resetting})), previously calculated by Chaudhuri et al.~\cite{Chaudhuri2021}, 
\bea
\la\rv^2(t)\ra &=& \f{2}{\beta}\left(1-e^{-2\beta t}\right) +\f{\Pe^2}{\beta(1+\beta)} - \f{2\Pe^2 e^{-\beta t}}{1-\beta} \left[\f{e^{-\beta t}}{2\beta} - \f{e^{-t}}{1+\beta}\right]\,.
\label{eq:r2avg_wo_resetting}
\eea
At small times ($t\to 0$), the expansion of equation~(\ref{eq:r2avg_wo_resetting}) leads to
\bea
\la\rv^2\ra &\simeq& 4t + (\Pe^2 -4\beta) t^2 + \f{1}{3}\left[8\beta^2 - \Pe^2 - 3 \beta \Pe^2\right]t^3+\mathcal{O}(t^4)\,.
\eea
In the steady state ($t\to \infty$), equation~(\ref{eq:r2avg_wo_resetting}) gives
\bea
\la\rv^2\ra^{st} &=& \f{2}{\beta}+\f{\Pe^2}{\beta(1+\beta)}\,.
\label{eq:MSD_st_ABP_wo_resetting}
\eea


\noindent
The effective diffusion coefficient $D_{\rm eff} = \beta \la \rv^2 \ra_{\rm st}/2$
\bea
D_{\rm eff} = 1 + \f{\Pe^2}{2(1+\beta)}\,,
\eea
which enhanced with increase of activity $\Pe$ and suppressed with increase of  harmonic stiffness $\beta$.

\subsection{ABP under complete stochastic resetting}
In the absence of harmonic trap ($\beta=0$), equation~(\ref{eq:r2avg_resetting}) simplifies to
\bea
\la\rv^2(t)\ra_r &=& \f{4(1+r)+2\Pe^2}{r(1+r)}-\f{(4+2\Pe^2)e^{-rt}}{r} + \f{2\Pe^2e^{-(1+r)t}}{(1+r)}\,. 
\label{eq:r2avg_ABP_resetting_wo_ht}
\eea
At small times ($t\to 0$), the expansion of equation~(\ref{eq:r2avg_ABP_resetting_wo_ht}) leads to
\bea
\la\rv^2(t)\ra_r &\simeq& 4t + (\Pe^2-2r) t^2 - \f{1}{3}\left[\Pe^2(1+2r)-2r^2\right]t^3+\mathcal{O}(t^4)\,.
\eea
The MSD at small time exhibits diffusive behavior $\la\rv^2\ra_r=4t$, which crosses over to ballistic behavior $\la\rv^2\ra_r\sim t^2$ when $\Pe^2>2r$ at $t^{*}=4/(\Pe^2-2r)$. At long times ($t\to\infty$), $\la\rv^2\ra_r$ reaches a steady state,
\bea
\la\rv^2\ra^{st}_r &=& \f{4}{r} + \f{2 \Pe^2}{r(1+r)}~.
\label{eq:MSD_st_ABP_wo_ht}
\eea
It reaches steady state at $t^{st}= \la\rv^2\ra^{st}_r /4$ for $\Pe^2\leq 2r$ and $t^{st}=[\la\rv^2\ra^{st}_r/(\Pe^2-2r)]^{1/2}$ for $\Pe^2>2r$. 

\noindent
{\em Note}, the steady state MSD in equations~(\ref{eq:MSD_st_ABP_wo_ht}) and (\ref{eq:MSD_st_ABP_wo_resetting}) have same form, with $\la\rv^2\ra^{st}_r = 2 \la\rv^2\ra^{st}$. Thus, in the steady state, the impact of stochastic resetting rate and trap stiffness on ABP is equivalent.

\noindent
The effective diffusion coefficient $D_{\rm eff} = r \la \rv^2 \ra_{\rm st}/2$
\bea
D_{\rm eff} = 2 \left[ 1 + \f{\Pe^2}{2(1+r)} \right]\,.
\eea
This is twice the effective diffusion coefficient of ABP in a harmonic trap. Similar to ABP in a harmonic trap, the effective diffusion coefficient here is enhanced with an increase in activity $\Pe$ and suppressed with increase of resetting rate $r$.

\subsection{Brownian particle under stochastic resetting in a harmonic trap}
In the absence of activity ($\Pe=0$), equation~(\ref{eq:r2avg_resetting}) simplifies to
\bea
\la\rv^2(t)\ra_r =  \f{4\left(1-e^{-(2\beta+r) t}\right)}{2\beta+r}\,.
\label{eq:r2avg_BP_resetting_w_ht}
\eea
At small times ($t\to 0$), the expansion of equation~(\ref{eq:r2avg_BP_resetting_w_ht}) leads to
\bea
\la\rv^2\ra_r \simeq 4t -2(2\beta + r) t^2 + \f{2}{3}\left[4\beta^2 + 4\beta r +r^2\right]t^3+\mathcal{O}(t^4)\,.
\eea
The MSD at small time exhibits diffusive behavior $\la\rv^2\ra_r=4t$, which reaches steady state ($t\to\infty$), equation~(\ref{eq:r2avg_BP_resetting_w_ht}) gives
\bea
\la\rv^2\ra^{st}_r=\f{4}{(2\beta+r)}\,.
\eea

\noindent
The effective diffusion coefficient $D_{\rm eff} = (\beta+r) \la \rv^2 \ra_{\rm st}/2$
\bea
D_{\rm eff} = \f{2(\beta+r)}{(2\beta+r)}\,.
\eea
The effective diffusion coefficient increases with increase of resetting rate $r$.

\subsection{Brownian particle under stochastic resetting}
In the absence of activity ($\Pe=0$) and harmonic trap ($\beta=0$), equation~(\ref{eq:r2avg_resetting}) simplifies to
\bea
\la\rv^2(t)\ra_r =  \f{4\left(1-e^{-r t}\right)}{r}\,.
\label{eq:r2avg_BP_resetting_wo_ht}
\eea
At small times ($t\to 0$), the expansion of equation~(\ref{eq:r2avg_BP_resetting_wo_ht}) leads to
\bea
\la\rv^2\ra_r \simeq 4t -2r t^2 + \f{2}{3} r^2 t^3+\mathcal{O}(t^4)\,.
\eea
In the steady state ($t\to\infty$), equation~(\ref{eq:r2avg_BP_resetting_wo_ht}) gives
\bea
\la\rv^2\ra^{st}_r=\f{4}{r}\,.
\eea

\noindent
The effective diffusion coefficient $D_{\rm eff} = r \la \rv^2 \ra_{\rm st}/2$
\bea
D_{\rm eff} = 2\,.
\eea
The effective diffusion coefficient is constant and twice of the thermal diffusion coefficient.

\section{Displacement fluctuations and its limiting cases}
\label{app:disp_fluct}

The displacement fluctuation $\la\d\rv^2\ra_r=\la\rv^2\ra_r-\la\rv\ra_{r}^2$.
At small time ($t\to 0$),
\bea
\la\d\rv^2\ra_r &\simeq& 4t -2 (2\beta +  r) t^2 +\f{2}{3}\left[4\beta^2 + \Pe^2 + 4\beta r +r^2\right]t^3+\mathcal{O}(t^4)\,.
\eea
In the steady state,
\bea
\la\d\rv^2\ra^{st}_r &=& \f{4[2+\Pe^2+3r+r^2+2\beta+\beta r]}{(2+r)(1+\beta+r)(2\beta+r)}\,.
\eea
The limiting cases are discussed below.
\subsection{ABP in a harmonic trap}
The displacement fluctuation $\la\d\rv^2\ra=\la\rv^2\ra-\la\rv\ra^2$. Here, we set resetting rate $r=0$.

\noindent
At small times ($t\to 0$),
\bea
\la\d\rv^2\ra &\simeq& 4t -4\beta t^2 + \f{2}{3}\left[4\beta^2 + \Pe^2\right]t^3+\mathcal{O}(t^4)\,.
\eea
In the steady state ($t\to\infty$),
\bea
\la\d\rv^2\ra^{st} &=& \f{2}{\beta}+\f{\Pe^2}{\beta(1+\beta)}\,.
\eea

\subsection{ABP under complete stochastic resetting}

The displacement fluctuation $\la\d\rv^2\ra_r=\la\rv^2\ra_r-\la\rv\ra_{r}^2$. Here, we set trap strength $\beta=0$.

\noindent
At small times ($t\to 0$),
\bea
\la\d\rv^2\ra_r &\simeq& 4t -2  r t^2 + \f{2}{3}\left(\Pe^2 +r^2\right)t^3+\mathcal{O}(t^4)\,.
\eea
In the steady state ($t\to\infty$),
\bea
\la\d\rv^2\ra^{st}_r &=& \f{4}{r} + \f{4\Pe^2}{r(1+r)(2+r)}\,.
\eea

\subsection{Brownian particle under stochastic resetting in a harmonic trap}
Here, we set activity $\Pe=0$.
The displacement fluctuation $\la\d\rv^2\ra_r=\la\rv^2\ra_r-\la\rv\ra_{r}^2=\la\rv^2\ra_r$.

\subsection{Brownian particle under stochastic resetting}
Here, we set activity $\Pe=0$ and trap strength $\beta=0$.
The displacement fluctuation $\la\d\rv^2\ra_r=\la\rv^2\ra_r-\la\rv\ra_{r}^2=\la\rv^2\ra_r$.

\section{Second order moment and fluctuations along initial orientation direction}
\label{app:second_moment_parallel}
\subsection{ABP in a harmonic trap}
The initial orientation of the ABP along $x$-direction $\uv_0=\hat{x}$. Using $\psi=\rpara^{2}=x^2$, we get using equation~(\ref{eq:moment}),
\bea
(s + 2\beta) \la\rpara^2\ra_s &=&  2/s + 2 \Pe \la x u_x\ra_s\,.
\eea
Now, we proceed to calculate the second term $\la x u_x \ra_s$ assuming $\psi=x u_x$, we get using equation~(\ref{eq:moment}), 
\bea
\la x u_x\ra &=& \f{\Pe \la u_x^2\ra_s}{s+1+\beta}\,. 
\eea
Again, we need to calculate $\la u_x^2\ra_s$, substituting $\psi=u_x^2$ in equation~(\ref{eq:moment}), we get
\bea
\la u_x^2\ra_s &=& \f{s+2}{s(s+4)}\,.
\eea

Finally,

\bea
 \la\rpara^2\ra_s &=& \f{1}{(s + 2\beta)} \left[\f{2}{s} + \f{2\Pe^2 (s+2)}{s(s+4)(s+1+\beta)}  \right]\,.
\eea
Inverse Laplace transformation of the above equation leads to the $\la\rpara^2(t)\ra$ for ABP in a harmonic trap in the absence of stochastic resetting
\bea
\fl\la\rpara^2(t)\ra = \frac{e^{-4t} \Pe^2}{2 (-3 + \beta) (-2 + \beta)} - 
\frac{2 e^{-(1 + \beta)t} \Pe^2}{(-3 + \beta) (1 + \beta)} + 
\frac{e^{-2\beta t} (2 - \beta + \Pe^2)}{(-2 + \beta) \beta} + 
\frac{2 + 2\beta + \Pe^2}{2 \beta (1 + \beta)}\,. \nonumber\\
\eea

\noindent
In the small time limit ($t\to 0$), $\la \rpara^2(t)\ra$ gives
\bea
&\la \rpara^2(t)\ra  \simeq  2t + (-2 \beta + \Pe^2) t^2 + 
\frac{1}{3} \left(4 \beta^2 - 3 \Pe^2 - 3 \beta \Pe^2\right) t^3 +\mathcal{O}(t^4)\,.
\eea

\noindent
In the steady state ($t\to\infty$), we get 
\bea
\la \rpara^2\ra^{st} &=& \frac{2(1+\beta)+\Pe^2}{2\beta(1+\beta)}\,.
\eea

\noindent
Displacement fluctuations along parallel direction $\la \delta \rpara^2\ra_r= \la \rpara^2\ra_r-\la \rpara\ra^{2}_r$.
\bea
\la\delta\rpara^2(t)\ra_r &=
\frac{1}{2} \left(\frac{e^{-4t} \Pe^2}{6 - 5 \beta + \beta^2} - \frac{4 e^{-(1 + \beta)t} \Pe^2}{(-3 + \beta) (1 + \beta)} - \frac{2 (e^{-t} - e^{-\beta t})^2 \Pe^2}{(-1 + \beta)^2} \right.\nn\\
&\left.+ \frac{2 e^{-2 \beta t} (2 - \beta + \Pe^2)}{(-2 + \beta) \beta} + \frac{2 + 2 \beta + \Pe^2}{\beta + \beta^2}\right)\,.
\eea

\noindent
In the small time limit ($t\to 0$), $\la \rpara^2(t)\ra_r$ gives
\bea
\la \d\rpara^2(t)\ra_{r} & \simeq & 2 t - 2 \beta t^2 +\f{4\beta^2 t^3}{3} + \mathcal{O}(t^4)\,.
\eea

\noindent
In the steady state ($t\to\infty$), we get 
\bea
\la \d\rpara^2\ra_{r}^{st} =  \la \rpara^2\ra^{st} = \frac{2(1+\beta)+\Pe^2}{2\beta(1+\beta)}\,.
\eea

\noindent
Now the perpendicular component of displacement fluctuation $\la \delta \rperp^2\ra_r= \la\d\rv^2\ra_r - \la\d\rpara^2\ra_r$.
\bea
\la\d\rperp^2(t)\ra_r &=
-\frac{e^{-4t} \Pe^2}{2 (6 - 5 \beta + \beta^2)} + \frac{4 e^{-(1 + \beta)t} \Pe^2}{3 - \beta - 3 \beta^2 + \beta^3} + \frac{2 + 2 \beta + \Pe^2}{2 \beta + 2 \beta^2} \nonumber\\
&- \frac{e^{-2 \beta t} (2 - 3 \beta + \beta^2 + \Pe^2)}{\beta (2 - 3 \beta + \beta^2)}\,.
\eea

In the small time limit ($t\to 0$), $\la \rpara^2(t)\ra_r$ gives
\bea
\fl\la \d\rperp^2(t)\ra  \simeq  2 t - 2 \beta t^2 +\f{2}{3}(2\beta^2+\Pe^2) t^3 - \f{1}{6} \left[4\beta^3 +5 \Pe^2 +3 \beta \Pe^2\right] t^4 + \mathcal{O}(t^5)\,.
\eea

In the steady state ($t\to\infty$), we get 
\bea
\la \d\rperp^2\ra_{r}^{st} = \la \delta \rpara^2\ra^{st}_r =  \la \rpara^2\ra^{st} = \frac{2(1+\beta)+\Pe^2}{2\beta(1+\beta)}\,.
\eea

\subsection{ABP under stochastic resetting in a harmonic trap}

Now, we get the $\la\rpara^2(t)\ra$ under stochastic position and orientation resetting using equation~(\ref{eq:moment_resetting})
\bea
\fl\la\rpara^2(t)\ra_r = e^{-rt} \left( 
\frac{e^{-4t} \Pe^2}{2 (-3 + \beta) (-2 + \beta)} - 
\frac{2 e^{-(1 + \beta)t} \Pe^2}{(-3 + \beta) (1 + \beta)} + 
\frac{e^{-2\beta t} (2 - \beta + \Pe^2)}{(-2 + \beta) \beta} + 
\frac{2 + 2\beta + \Pe^2}{2 \beta (1 + \beta)}
\right) \nonumber\\
\fl+  
\frac{\beta (2 - 2 e^{-rt}) - e^{-(2\beta + r)t} (-1 + e^{2\beta t}) r}{\beta (2 \beta + r)} + 
\frac{r \Pe^2}{2} \left[
-\frac{e^{-(4 + r)t}}{(6 - 5 \beta + \beta^2) (4 + r)} + 
\frac{4 e^{-(1 + \beta + r)t}}{(-3 + \beta) (1 + \beta) (1 + \beta + r)} \right.\nonumber\\
\fl\left.- 
\frac{2 e^{-(2 \beta + r)t}}{(-2 + \beta) \beta (2 \beta + r)} + 
\frac{4 (2 + r)}{r (4 + r) (1 + \beta + r) (2 \beta + r)} - 
\frac{e^{-rt}}{\beta r + \beta^2 r} 
\right]\,.
\eea
In the small time limit ($t\to 0$), $\la \rpara^2(t)\ra_r$ gives
\bea
\fl\la \rpara^2(t)\ra \simeq 
2t + (-2 \beta + \Pe^2 - r) t^2 + 
\frac{1}{3} \left(4 \beta^2 - 3 \Pe^2 - 3 \beta \Pe^2 + 4 \beta r - 2 \Pe^2 r + r^2\right) t^3\nonumber\\
\fl+\mathcal{O}(t^4)\,.
\eea
In the steady state ($t\to\infty$), we get 
\bea
\la \rpara^2\ra_{r}^{st} &=& \frac{2 [(1+r)(4+r) + \Pe^2 (2 + r) + \beta (4 + r)]}{(4 + r) (1 + \beta + r) (2 \beta + r)}\,.
\eea

\noindent
Displacement fluctuations along parallel direction $\la \delta \rpara^2\ra_r= \la \rpara^2\ra_r-\la \rpara\ra^{2}_r$.

\bea
\fl\la\delta\rpara^2(t)\ra_r = 
\frac{1}{2} \left[e^{-rt} \left(\frac{e^{-4t} \Pe^2}{6 - 5 \beta + \beta^2} - \frac{4 e^{-(1 + \beta)t} \Pe^2}{(-3 + \beta) (1 + \beta)} + \frac{2 e^{-2 \beta t} (2 - \beta + \Pe^2)}{(-2 + \beta) \beta} + \frac{2 + 2 \beta + \Pe^2}{\beta + \beta^2}\right)\right. \nn\\
\fl\left.- \frac{2 \Pe^2 (r - \beta r + \beta e^{-(\beta + r)t} (1 + r) - e^{-(1 + r)t} (\beta + r))^2}{(-1 + \beta)^2 (1 + r)^2 (\beta + r)^2} \right.\nn\\
\fl\left. + \Pe^2 r \left(-\frac{e^{-(4 + r)t}}{(6 - 5 \beta + \beta^2) (4 + r)} + \frac{4 e^{-(1 + \beta + r)t}}{(-3 + \beta) (1 + \beta) (1 + \beta + r)} - \frac{2 e^{-(2 \beta + r)t}}{(-2 + \beta) \beta (2 \beta + r)} \right.\right. \nn\\
\fl\left.\left. + \frac{4 (2 + r)}{r (4 + r) (1 + \beta + r) (2 \beta + r)} - \frac{e^{-rt}}{\beta r + \beta^2 r}\right)+ \frac{\beta (4 - 4 e^{-rt}) - 4 e^{-(\beta + r)t} r \sinh(\beta t)}{\beta (2 \beta + r)}\right]\,.
\eea
In the small time limit ($t\to 0$), $\la \rpara^2(t)\ra_r$ gives
\bea
\fl\la \d\rpara^2(t)\ra  \simeq  2t - (2 \beta + r) t^2 + \frac{1}{3} \left(4 \beta^2 + 4 \beta r + \Pe^2 r + r^2\right) t^3 + \mathcal{O}(t^4)\,.
\eea
In the steady state ($t\to\infty$), we get 
\bea
\fl\la \d\rpara^2\ra_{r}^{st} =
-\frac{\Pe^2 r^2}{(1 + r)^2 (\beta + r)^2} + \frac{2}{2 \beta + r} + 
\frac{2 \Pe^2 (2 + r)}{(4 + r) (1 + \beta + r) (2 \beta + r)}\,.
\eea

\subsection{ABP under stochastic resetting}
\bea
\fl\la\rpara^2(t)\ra_r =
\frac{e^{-(4 + r)t}}{3r (1 + r) (4 + r)} \left(\Pe^2 r (1 + r) + 
2 e^{3t} \Pe^2 r (4 + r) - 3 e^{4t} (2 + \Pe^2) (1 + r) (4 + r) \right)\nn\\
\fl + \frac{2[\Pe^2 (2 + r) + (1 + r) (4 + r)]}{r (1 + r) (4 + r)}\,.  
\eea
In the small time limit ($t\to 0$), $\la \rpara^2(t)\ra_r$ gives
\bea
\la \rpara^2(t)\ra & \simeq &
2t + ( \Pe^2 - r) t^2 + 
\frac{1}{3} \left(- 3 \Pe^2 - 2 \Pe^2 r + r^2\right) t^3 +\mathcal{O}(t^4)\,.
\eea
In the steady state ($t\to\infty$), we get 
\bea
\la \rpara^2\ra_{r}^{st} &=& \frac{2[\Pe^2 (2 + r) + (1 + r) (4 + r)]}{r (1 + r) (4 + r) }\,.
\eea
Displacement fluctuations along parallel direction $\la \delta \rpara^2\ra_r= \la \rpara^2\ra_r-\la \rpara\ra^{2}_r$.
\bea
\fl\la\delta\rpara^2(t)\ra_r =
-\frac{e^{-2 (1 + r)t} (-1 + e^{(1 + r)t})^2 \Pe^2}{(1 + r)^2} + 
e^{-rt} \left(2t + \frac{1}{12} \Pe^2 \left(-9 + e^{-4t} + 8 e^{-t} + 12t\right)\right) \nn\\
\fl+ \frac{e^{-(4 + r)t}}{12r (1 + r) (4 + r)} \left[-\Pe^2 r^2 (1 + r) - 8 e^{3t} \Pe^2 r^2 (4 + r) \right.\nn\\
\fl\left.+ 
24 e^{(4 + r)t} (\Pe^2 (2 + r) + (1 + r) (4 + r))- 
3 e^{4t} (1 + r) (4 + r) (8 + 8rt + \Pe^2 (4 - 3r + 4rt))\right]\,.\nonumber\\
\eea
In the small time limit ($t\to 0$), $\la \rpara^2(t)\ra_r$ gives
\bea
\la \d\rpara^2(t)\ra & \simeq &
2t - r t^2 + 
\frac{r}{3} \left(\Pe^2 + r\right) t^3 +\mathcal{O}(t^4)\,.
\eea
In the steady state ($t\to\infty$), we get 
\bea
\la \d\rpara^2\ra_{r}^{st} &=& \frac{2(1+r)^2(4+r)+\Pe^2(4+2r+r^2)}{r (1 + r)^2 (4 + r) }\,.
\eea

\subsection{Brownian particle under stochastic resetting in a harmonic trap}

In the absence of activity ($\Pe=0$), simplifies to Brownian particle under stochastic resetting
\bea
\la\rpara^2(t)\ra_r &=& \f{2}{2\beta+r} \left(1-e^{-(2\beta+r)t}\right)\,.
\eea
In the small time limit ($t\to 0$), $\la \rpara^2(t)\ra_r$ gives the expansion for Brownian particle under stochastic resetting
\bea
\la \rpara^2(t)\ra_r & \simeq & 2t - (2 \beta  + r) t^2 + 
\frac{1}{3} \left(4 \beta^2 + 4 \beta r + r^2\right) t^3 +\mathcal{O}(t^4)\,.
\eea
In the steady state ($t\to\infty$), we get the expression for Brownian particle under stochastic resetting
\bea
\la \rpara^2\ra^{st}_r &=& \frac{2}{2\beta+r}\,.
\eea
Displacement fluctuations along parallel direction $\la \delta \rpara^2\ra_r= \la \rpara^2\ra_r-\la \rpara\ra^{2}_r=\la \rpara^2\ra_r$.

\subsection{Brownian particle under stochastic resetting}

In the absence of activity ($\Pe=0$), simplifies to Brownian particle under stochastic resetting
\bea
\la\rpara^2(t)\ra_r &=& \f{2}{r} \left(1-e^{-rt}\right)\,.
\eea
In the small time limit ($t\to 0$), $\la \rpara^2(t)\ra_r$ gives the expansion for Brownian particle under stochastic resetting
\bea
\la \rpara^2(t)\ra_r & \simeq & 2t -  r t^2 + 
\frac{r^2t^3}{3} +\mathcal{O}(t^4)\,.
\eea
In the steady state ($t\to\infty$), we get the expression for Brownian particle under stochastic resetting
\bea
\la \rpara^2\ra^{st}_r &=& \frac{2}{r}\,.
\eea
Displacement fluctuations along parallel direction $\la \delta \rpara^2\ra_r= \la \rpara^2\ra_r-\la \rpara\ra^{2}_r=\la \rpara^2\ra_r$.

\section{Detailed derivation of fourth order moment of displacement}
\label{app1}
Here, we show the detailed derivation of fourth moment of displacement under stochastic position and orientation resetting.

We get the fourth order moment of displacement in Laplace space $\la\rv^4\ra_s$ in the absence of stochastic resetting ($r=0$) with initial position at origin $\rv_0=0$ using equation~(\ref{eq:moment}), gives
\bea
\la\rv^4\ra_s &=& \f{1}{s+4\beta}\left[16\la\rv^2\ra_s + 4 \Pe \la(\uv\cdot\rv)\rv^2\ra_s\right]\,.
\label{eq:r4avg_Laplace}
\eea
Further, we proceed to calculate $\la(\uv\cdot\rv)\rv^2\ra_s$, 
\bea
\la(\uv\cdot\rv)\rv^2\ra_s &=& \f{\left[ 8\la\uv\cdot\rv\ra_s+\Pe\la\rv^2\ra_s+2\Pe\la(\uv\cdot\rv)^2\ra_s\right]}{(s+1+3\beta)}\,.
\eea
The first two quantities $\la\uv\cdot\rv\ra_s$ and $\la\rv^2\ra_s$ already calculated. The third term $\la(\uv\cdot\rv)^2\ra_s$ calculated as
\bea
\la(\uv\cdot\rv)^2\ra_s &=& \f{\left[2/s+2\la\rv^2\ra_s+2 \Pe\la\uv\cdot\rv\ra_s\right]}{(s+4+2\beta)}\,.
\eea
Finally, we get $\la\rv^4\ra_s$ in Laplace space after substituting $\la\rv^2\ra_s$ and $\la(\uv\cdot\rv)^2\ra_s$ in equation~(\ref{eq:r4avg_Laplace}). Inverse Laplace transformation  of  equation~(\ref{eq:r4avg_Laplace}) leads to $\la\rv^4(t)\ra$ for ABP in a harmonic trap without stochastic resetting already explored in Chaudhuri et al.~\cite{Chaudhuri2021}
\bea
\fl\la\rv^4(t)\ra = \frac{2\mathrm{Pe}^4 e^{-2 (2 + \beta) t} }{(-3 + \beta) (-2 + \beta) (2 + \beta) (3 + \beta)} - \frac{4 e^{-2 \beta t} [-4 + 4 \beta^2 - 4 \mathrm{Pe}^2 - \mathrm{Pe}^4]}{(-1 + \beta) \beta^2 (1 + \beta)} \nn\\
\fl+ \frac{4 e^{-(1 + 3 \beta) t} [-12 \mathrm{Pe}^2 - 32 \beta \mathrm{Pe}^2 + 12 \beta^2 \mathrm{Pe}^2 + 5 \mathrm{Pe}^4 - 3 \beta \mathrm{Pe}^4]}{(-3 + \beta) (-1 + \beta) \beta (1 + \beta) (1 + 3 \beta)} \nn\\
\fl+ \frac{e^{-4 \beta t} [-16 + 72 \beta - 80 \beta^2 + 24 \beta^3 - 16 \mathrm{Pe}^2 + 56 \beta \mathrm{Pe}^2 - 24 \beta^2 \mathrm{Pe}^2 - 4 \mathrm{Pe}^4 + 3 \beta \mathrm{Pe}^4]}{(-2 + \beta) (-1 + \beta) \beta^2 (-1 + 3 \beta)} \nn\\
\fl+ \frac{16 + 72 \beta + 80 \beta^2 + 24 \beta^3 + 16 \mathrm{Pe}^2 + 56 \beta \mathrm{Pe}^2 + 24 \beta^2 \mathrm{Pe}^2 + 4 \mathrm{Pe}^4 + 3 \beta \mathrm{Pe}^4}{\beta^2 (1 + \beta) (2 + \beta) (1 + 3 \beta)} \nn\\
\fl- \frac{4 e^{-(1 + \beta) t} [-12 \mathrm{Pe}^2 + 32 \beta \mathrm{Pe}^2 + 12 \beta^2 \mathrm{Pe}^2 + 5 \mathrm{Pe}^4 + 3 \beta \mathrm{Pe}^4]}{(-1 + \beta) \beta (1 + \beta) (3 + \beta) (-1 + 3 \beta)}\,.
\label{eq:r4avg}
\eea
In presence of resetting, substituting $\psi=\rv^4$ in equation~(\ref{eq:moment_resetting}) and using equation~(\ref{eq:r4avg}), we get
\bea
\fl\la\rv^4(t)\ra_r = \la\rv^4(t)\ra e^{-r t} +
 \frac{24 \beta^3 + 4 (2 + \mathrm{Pe}^2)^2 + 8 \beta^2 (10 + 3 \mathrm{Pe}^2) + \beta (72 + 56 \mathrm{Pe}^2 + 3 \mathrm{Pe}^4)}{\beta^2 (1 + \beta) (2 + \beta) (1 + 3 \beta) }\nn\\
\fl- \frac{4 r\mathrm{Pe}^2 [-12 + 12 \beta^2 + 5 \mathrm{Pe}^2 + \beta (32 + 3 \mathrm{Pe}^2)]}{(-1 + \beta) \beta (1 + \beta) (3 + \beta) (-1 + 3 \beta) (1 + \beta + r)} + \frac{4r [-4 \beta^2 + (2 + \mathrm{Pe}^2)^2]}{(-1 + \beta) \beta^2 (1 + \beta) (2 \beta + r)} \nn\\
\fl+ \frac{2 r\mathrm{Pe}^4}{(36 - 13 \beta^2 + \beta^4) (4 + 2 \beta + r)} + \frac{4 r\mathrm{Pe}^2 [-12 + 12 \beta^2 + 5 \mathrm{Pe}^2 - \beta (32 + 3 \mathrm{Pe}^2)]}{(-3 + \beta) (-1 + \beta) \beta (1 + \beta) (1 + 3 \beta) (1 + 3 \beta + r)} \nn\\
\fl+ \frac{24 r\beta^3 - 4 (2 + \mathrm{Pe}^2)^2 - 8 \beta^2 (10 + 3 \mathrm{Pe}^2) + \beta (72 + 56 \mathrm{Pe}^2 + 3 \mathrm{Pe}^4)}{(-2 + \beta) (-1 + \beta) \beta^2 (-1 + 3 \beta) (4 \beta + r)} \nn\\
\fl-\frac{[24 \beta^3 + 4 (2 + \mathrm{Pe}^2)^2 + 8 \beta^2 (10 + 3 \mathrm{Pe}^2) + \beta (72 + 56 \mathrm{Pe}^2 + 3 \mathrm{Pe}^4)]e^{-r t}}{\beta^2 (1 + \beta) (2 + \beta) (1 + 3 \beta)} \nn\\
\fl+ \frac{4 r \mathrm{Pe}^2 e^{-(1 + r + \beta) t} [-12 + 12 \beta^2 + 5 \mathrm{Pe}^2 + \beta (32 + 3 \mathrm{Pe}^2)]}{(-1 + \beta) \beta (1 + \beta) (3 + \beta) (-1 + 3 \beta) (1 + \beta + r)} \nn\\
\fl+ \frac{4 r e^{-(r+2 \beta) t} [4 \beta^2 - (2 + \mathrm{Pe}^2)^2]}{(-1 + \beta) \beta^2 (1 + \beta) (2 \beta + r)} - \frac{2 r \mathrm{Pe}^4 e^{-[2 (2 + \beta) +r]t}}{(36 - 13 \beta^2 + \beta^4) (4 + 2 \beta + r)} \nn\\
\fl+ \frac{4 r \mathrm{Pe}^2 e^{-(1 + r+ 3 \beta) t} [12 - 12 \beta^2 - 5 \mathrm{Pe}^2 + \beta (32 + 3 \mathrm{Pe}^2)]}{(-3 + \beta) (-1 + \beta) \beta (1 + \beta) (1 + 3 \beta) (1 + 3 \beta + r)} \nn\\
\fl+ \frac{re^{-(r+4 \beta) t} [-24 \beta^3 + 4 (2 + \mathrm{Pe}^2)^2 + 8 \beta^2 (10 + 3 \mathrm{Pe}^2) - \beta (72 + 56 \mathrm{Pe}^2 + 3 \mathrm{Pe}^4)]}{(-2 + \beta) (-1 + \beta) \beta^2 (-1 + 3 \beta) (4 \beta + r)}\,. 
\label{eq:r4avg_resetting}
\eea
At small time ($t\to 0$),
\bea
\fl\la\rv^4\ra_r \simeq 32 t^2 +\f{16}{3} \left(3\Pe^2-4r-12\beta\right)t^3 \nonumber\\
\fl+ \f{1}{3}\left[224 \beta^2+3 \Pe^4+24 r^2 +144\beta r -96 \beta \Pe^2 -16 \Pe^2-36r \Pe^2\right]t^4+\mathcal{O}(t^5)\,.
\label{eq:r4avg_resetting_small_time}
\eea
In the absence of activity by substituting $\Pe=0$ in equation~(\ref{eq:r4avg_resetting_small_time}), we get small time behavior of a Brownian particle in two dimensions under resetting
\bea
\la\rv^4\ra_{r} &\simeq& 2 t^2 -\f{64\left(r+3\beta\right)t^3}{3}  + \f{\left[224 \beta^2+24 r^2 +144\beta r\right]t^4}{3} +\mathcal{O}(t^5)\,.
\eea
In the steady state($t\to\infty$),
\bea
\fl\la\rv^4\ra^{st}_r = \frac{8}
{(1 + \beta + r) (2 \beta + r) (4 + 2 \beta + r) (1 + 3 \beta + r) (4 \beta + r)}\times\nonumber\\
\fl\left[48 \beta^3 + 8 (1 + r)^2 (4 + r) + \Pe^4 (8 + 3 r) + 8 \beta^2 (20 + 6 \Pe^2 + 11 r) \right.\nonumber\\
\fl \left.+ 4 \Pe^2 (8 + 14 r + 3 r^2) + 
     2 \beta \left[3 \Pe^4 + 8 \Pe^2 (7 + 3 r) + 24 (3 + 4 r + r^2)\right]\right]\,.
\label{eq:r4avg_resetting_st}     
\eea
In the absence of activity ($\Pe=0$), Brownian particle in two dimensions under resetting in a harmonic trap
\bea
\fl\la\rv^4\ra^{st}_r = \f{8\left[48\beta^3+8(1+r)^2(4+r)+8\beta^2(20+11r)+48\beta (3+4r+r^2)\right]}{(1+\beta+r)(2\beta+r)(4+2\beta+r)(1+3\beta+r)(4\beta+r)}\,.
\eea
In the absence of activity ($\Pe=0$) and harmonic trap ($\beta=0$), Brownian particle in two dimensions under resetting, $\la\rv^4\ra^{st}_r = 64/r^2$.

\section{Limiting cases of excess kurtosis}
\label{app:excess_kurtosis}

At small times ($t\to 0$), the excess kurtosis in equation~(\ref{eq:kurtosis}) results

\bea
\fl\mathcal{K}_r \simeq \f{rt}{3} - \f{(3\Pe^4+16\beta r-4r \Pe^2)}{96} t^2 + \f{1}{2880} \left[45 \Pe^6+60 \Pe^4 +32 r^2\Pe^2 +64 \beta^2 r \right.\nn\\
\fl\left.+24 \beta r \Pe^2-42r \Pe^4 -56 r\Pe^2-32r^3-128\beta r^2\right] t^3 + \mathcal{O}(t^4)\,.
\label{eq:kurtosis_small_time}
\eea

In the absence of activity ($\Pe=0$), simplifies to Brownian particle under stochastic resetting in two dimensions, 
\bea
\fl\mathcal{K}_r \simeq \f{rt}{3} - \f{\beta r t^2}{6} + \f{r\left[2\beta^2 -r^2-4\beta r\right]}{90} t^3 + \f{\beta r \left[ 4 \beta^2 + 12 \beta r + 3 r^2\right]}{360} + \mathcal{O}(t^5)\,.
\eea

In the absence of resetting rate ($r=0$), simplifies to ABP in a harmonic trap in two dimensions,
\bea
\fl\mathcal{K}_r \simeq -\f{\Pe^4 t^2}{32} + \f{\Pe^4 (4+3\Pe^2) t^3}{192} - \f{\Pe^4\left[ 136 - 120 \beta^2 +360 \Pe^2 +135 \Pe^4 \right]}{23040} t^4 + \mathcal{O}(t^5)\,.\nonumber\\
\eea

\bea
\fl\mathcal{K}^{st}_r (\Pe\to 0) \simeq \f{r}{r+4\beta} + \f{\Pe^2 (r^2+2\beta r)}{(1+\beta+r)(1+3\beta+r)(4\beta+r)} + \mathcal{O}(\Pe^3)\,.
\label{eq:kurtosis_st_Pe_to_0}
\eea

\section{ABP under stochastic resetting without harmonic trap}
\label{app:abps_stochastic_resetting_wo_ht}

\begin{figure*}[!t]
\begin{center}
\includegraphics[width=17cm]{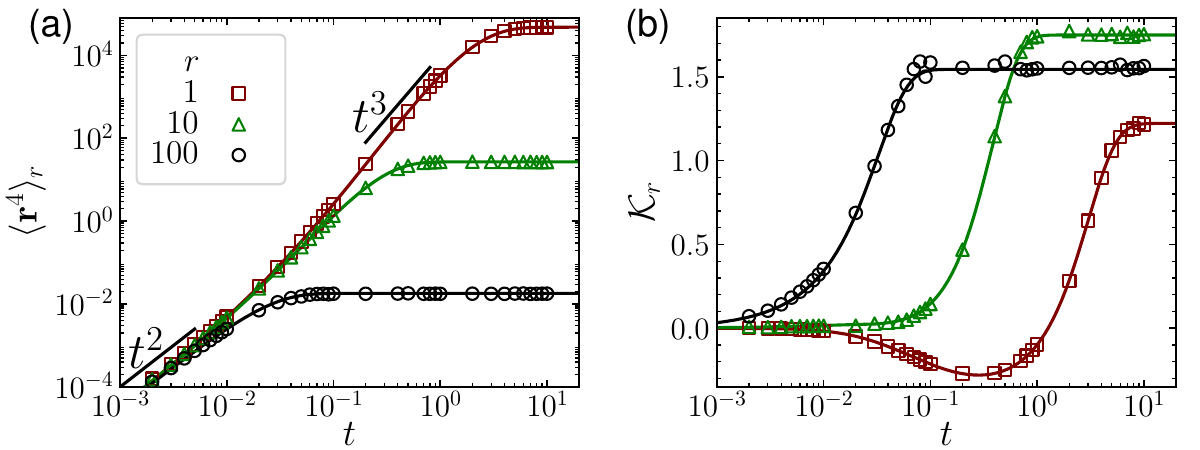} 
\caption{ABP under stochastic resetting without harmonic trap ($\beta=0$).
(a) Fourth order moment of displacement $\la\rv^4\ra_{r}$ (equation~(\ref{eq:r4avg_resetting})) and (b) Excess kurtosis $\mathcal{K}_r$ as a function of time $t$ for varying resetting rates $r=1 (\mathrm{squares}, \Box), 10 (\mathrm{triangles}, \triangle), 100 (\mathrm{circles},\circ)$ with $\Pe=10$. The solids lines are analytic predictions and symbols are from numerical simulations. The initial and resetting position is at the origin with the initial and resetting orientation along the $x$-axis.
} 
\label{app_fig2}
\end{center}
\end{figure*} 
Figure~\ref{app_fig2} presents the fourth order displacement moment in (a) and the excess kurtosis in (b) as a function of time of ABP under stochastic resetting without a harmonic trap ($\beta=0$).
In figure~\ref{app_fig2}(a), the analytic fourth order moment of displacement, $\la\rv^4\ra_{r}$, is shown as solid lines, compared to simulation results (points) for resetting rates $r=1,~10,~100$ with $\Pe=10$. For low resetting rates ($r=1$), ballistic behavior $\la\rv^4\ra_{r} \sim t^3$ is observed in the intermediate time regime.
In figure~\ref{app_fig2}(b), the analytic excess kurtosis, $\mathcal{K}_r$, is shown as solid lines, compared to simulation results (points) for the same resetting rates. Negative excess kurtosis is observed in the intermediate time regime for $r=1$.

\begin{figure*}[!t]
\begin{center}
\includegraphics[width=7.0cm]{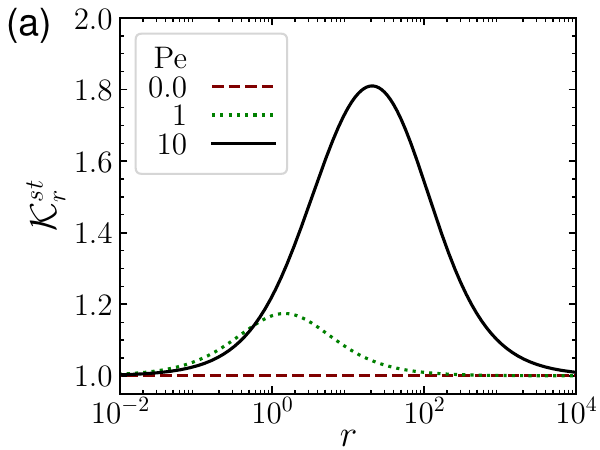}
\includegraphics[width=7.0cm]{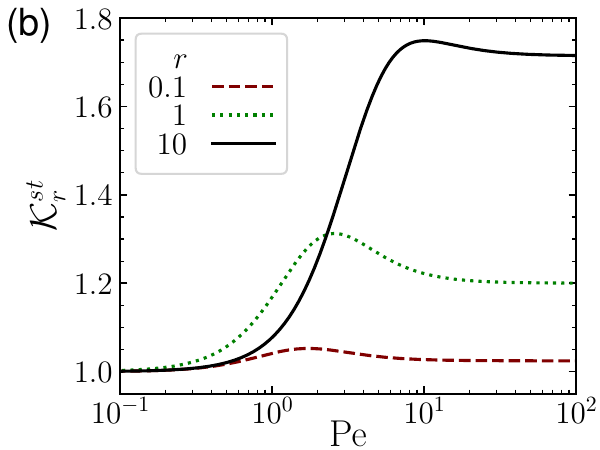}
\includegraphics[width=14cm]{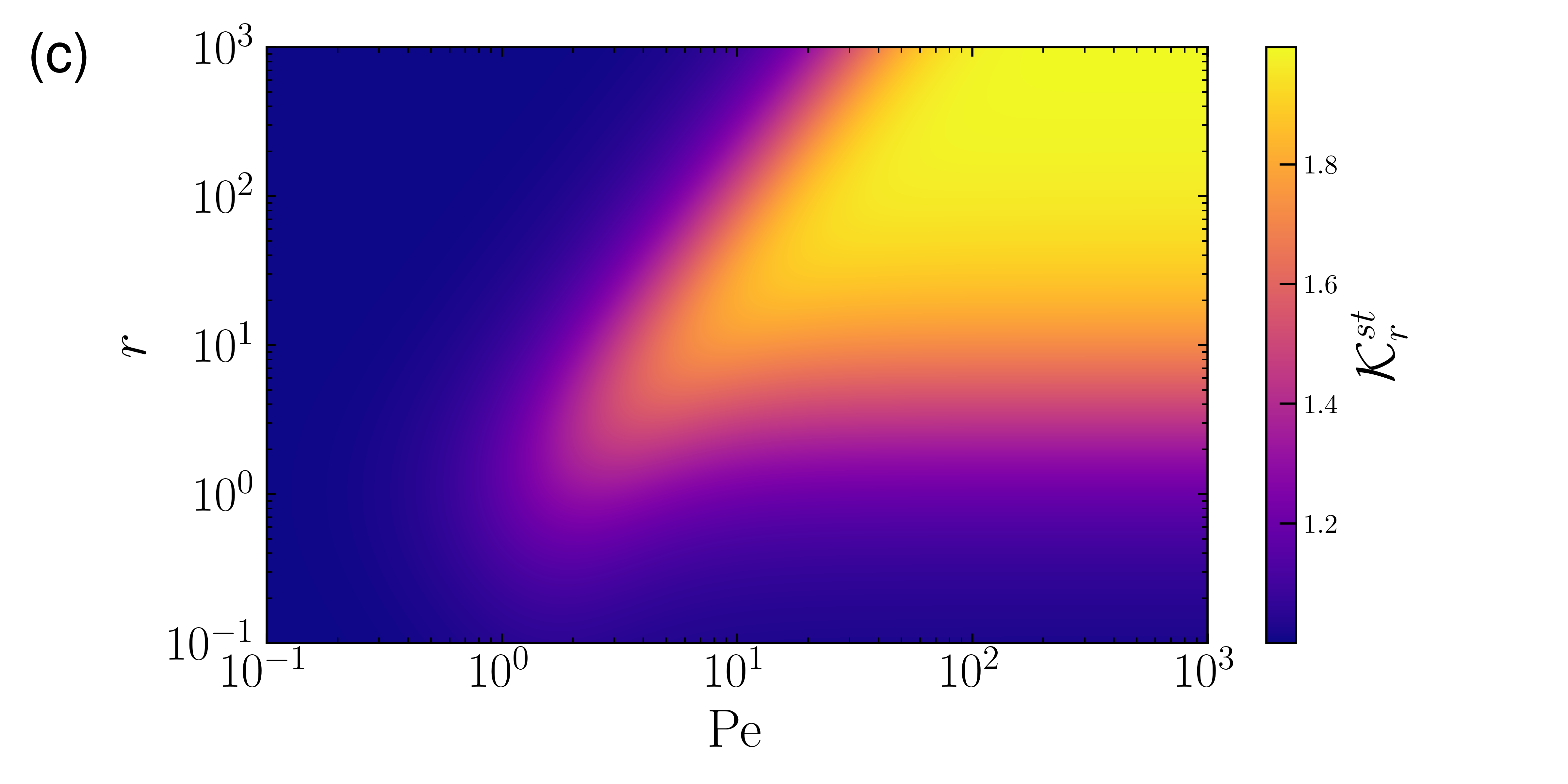}  
\caption{Steady state behavior of ABP under stochastic resetting without harmonic trap ($\beta=0$). Steady state excess kurtosis $\mathcal{K}^{st}_{r}$ (equation~(\ref{eq:kurtosis_st_wo_ht})) as a function of (a) resetting rates ($r$) for  activities $\Pe=0.1,~1,~10$ and as a function of (b) $\Pe$ for $r=0.1,~1,~10$. (c) Plot of $\mathcal{K}^{st}_{r}$ in $r-\Pe$ plane.} 
\label{app_fig3}
\end{center}
\end{figure*} 
Figure~\ref{app_fig3} presents the steady state quantification of ABP under stochastic resetting without a harmonic trap ($\beta=0$).
In figure~\ref{app_fig3}(a), we show steady state excess kurtosis $\mathcal{K}^{st}_r$ (equation~(\ref{eq:kurtosis_st_wo_ht})) as a function of $r$ for $\Pe=0.1,~1,~10$. 
In figure~\ref{app_fig3}(b), we show the $\mathcal{K}^{st}_r$ (equation~(\ref{eq:kurtosis_st_wo_ht})) as a function of $\Pe$ for $r=0.1,~1,~10$. We see the non-monotonic behavior of $\mathcal{K}^{st}_r$ as a function of $r$ with high activity $\Pe$ (figure~\ref{app_fig3}(a)). It demonstrates the presence of an optimal resetting rate where $\mathcal{K}^{st}_r$ is maximized for fixed activity, suggesting a maximum heavy tail in the position distribution, which increases the likelihood of finding the particle at a very long distance. We can also observe non-monotonic in $\mathcal{K}^{st}_r$ as a function of $\Pe$ for constant $r$, which peaks at intermediate $\Pe$ (figure~\ref{app_fig3}(b)). In figure~\ref{app_fig3}(c), we show the $\mathcal{K}^{st}_r$ in $r-\Pe$ plane, exhibits this non-monotonic behavior in both parameters $r$ and $\Pe$.

We found that at intermediate timescales, high activity ($\Pe$) and low resetting rate ($r$) result in persistent behavior, characterized by bimodal distributions and negative excess kurtosis. At longer times, the system reaches a nonequilibrium steady state (NESS), where moments stabilize. The steady state excess kurtosis shows non-monotonic behavior, deviating from the positive excess kurtosis value of $1$ seen in Brownian particle under resetting, across different values of both activity ($\Pe$) and resetting rate ($r$).

\section*{References}
\bibliographystyle{iopart-num}
\bibliography{reference}

\end{document}